\documentclass[a4paper,fleqn,usenatbib]{mnras}

\usepackage{newtxtext,newtxmath}
\usepackage[T1]{fontenc}
\usepackage{ae,aecompl}

\usepackage{graphicx}	% Including figure files
\usepackage{amsmath}	% Advanced maths commands
\title[The origin of globular clusters]
      {On the origin of globular clusters in a hierarchical universe} 

\author[G.~De Lucia et al.]
       {Gabriella De Lucia$^{1,2}$\thanks{Email: gabriella.delucia@inaf.it},
         J.~M.~Diederik Kruijssen$^{3,4,5}$,
         Sebastian Trujillo-Gomez$^{6}$, \newauthor Michaela Hirschmann$^{7,1}$, Lizhi Xie$^8$\\
        $^1$INAF - Astronomical Observatory of Trieste, via G.B. Tiepolo 11, 
        I-34143 Trieste, Italy\\
        $^2$IFPU - Institute for Fundamental Physics of the Universe, via Beirut 2, 34151, Trieste, Italy\\
        $^3$Technical University of Munich, School of Engineering and Design, Department of Aerospace and Geodesy, Chair of Remote Sensing Technology, \\\hspace{2.2mm}Arcisstr. 21, 80333 Munich, Germany\\
        $^4$Cosmic Origins Of Life (COOL) Research DAO, coolresearch.io\\
        $^5$Max-Planck Institut f\"ur Astrophysik, Karl-Schwarzschild-Stra\ss e 1, D-85748 Garching, Germany\\
        $^6$Astroinformatics Group, Heidelberg Institute for Theoretical Studies, Schloss-Wolfsbrunnenweg 35, 69118 Heidelberg, Germany\\
        $^7$Institute for Physics, Laboratory for Galaxy Evolution, EPFL, Observatoire de Sauverny, Chemin Pegasi 51, 1290 Versoix, Switzerland\\
        $^8$Tianjin Normal University, Binshuixidao 393, 300387, Tianjin, People’s Republic of China}

\date{Accepted XXX. Received YYY; in original form ZZZ}

\pubyear{2023}

\begin{document}
\label{firstpage}
\pagerange{\pageref{firstpage}--\pageref{lastpage}}
\maketitle

\begin{abstract}
  We present an end-to-end description of the formation of globular clusters (GCs) combining a treatment for their formation and dynamical evolution within galaxy haloes with a state-of-the-art semi-analytic simulation of galaxy formation. Our approach allows us to obtain exquisite statistics to study the effect of the environment and assembly history of galaxies, while still allowing a very efficient exploration of the parameter space. Our reference model, including both efficient cluster disruption during galaxy mergers and dynamical friction of GCs within the galactic potential, accurately reproduces the observed correlation between the total mass in GCs and the parent halo mass. A deviation from linearity is predicted at low halo masses, which is driven by a strong dependence on morphological type: bulge-dominated galaxies tend to host larger masses of GCs than their later-type counterparts. While the significance of the difference might be affected by resolution at the lowest halo masses considered, this is a robust prediction of our model and  a natural consequence of the assumption that cluster migration into the halo is triggered by galaxy mergers. Our model requires an environmental dependence of GC radii to reproduce the observed low-mass mass distribution of GCs in our Galaxy. At GC masses $>10^6\,{\rm M}_\odot$, our model predicts fewer GCs than observed, due to an overly aggressive treatment of dynamical friction. Our model reproduces well the metallicity distribution measured for Galactic GCs, even though we predict systematically younger GCs than observed. We argue that this adds further evidence for an anomalously early formation of the stars in our Galaxy.
\end{abstract}


%%%%%%%%%%%%%%%%%%%%%%%%%%%%%%%%%%%%%%%%%%%%%%%%%%%%%%%%%%%%%%%%%%%%%%%%%%%%%%%
\section{Introduction}
\label{sec:intro}

Globular clusters (GCs) are found in all galaxies in the local Universe down
to galaxy stellar masses of $\sim 10^8\,{\rm M}_{\odot}$.  GCs typically have old ages ($\sim10$~Gyr,
\citealt{Strader_etal_2005,Forbes_etal_2010,VandenBerg_etal_2013}), nearly
uniform sizes \citep[e.g.][]{Masters_etal_2010,Krumholz_etal_2019}, and a peaked mass distribution that
can be approximated by a log-normal function, with a characteristic peak mass
that depends weakly on the host galaxy stellar mass
\citep[e.g.][]{Jordan_etal_2007}.  The old ages and small sizes of GCs have long prevented
direct observations of their formation - a situation that is changing rapidly with the new James Webb Telescope finally in operation \citep[see e.g.][]{Mowla_etal_2022,Vanzella_etal_2022,Claeyssens_etal_2023} %with available instrumentation, but this will eventually become feasible in the future: taking advantage of the strong magnification lensing signal of massive clusters imaged with the Hubble Space Telescope, and of spectroscopic follow-up with VLT/MUSE, \citet{Vanzella_etal_2017} studied the properties of five compact and very young objects at $z=2-6$. Given the small sizes (estimated half-light radii are between 16 and 140 pc) and the very young estimated ages ($<10$~Myr), \citet{Vanzella_etal_2017} argued that these objects might correspond to GCs in formation. Upcoming studies with JWST will reveal the true nature of these faint high-redshift systems.
    
For decades, the origin
of GCs has represented a largely unsolved problem that encompasses the fields of
star and galaxy formation. First theoretical work on this subject envisioned
that GC formation could be triggered by special conditions in the early
Universe: e.g.\ \citet{Peebles_and_Dicke_1968} argued that GCs may have formed
before the first galaxies, with masses determined by the Jeans mass. In a later
work by \citet{Fall_and_Rees_1985}, GCs were assumed to form during the
collapse of protogalaxies due to thermal instabilities in the hot gaseous
haloes. Alternative models pushed for a significantly later formation of GCs,
possibly triggered by mergers between gas-rich disk galaxies that can compress
and shock the interstellar medium \citep[ISM;][]{Schweizer_1987,Ashman_and_Zepf_1992}. At present, none of these scenarios are thought to explain the origin of the majority of GCs (see \citealt{Kruijssen_2014} and \citealt{Forbes_etal_2018} for reviews).

Important information about the physical processes leading to GC
formation can be inferred from their present day properties and from the
formation of (young) massive stellar clusters in the local Universe. A key insight has been that young GCs are observed to form in the local Universe whenever conditions are present that mimic those in high-redshift galaxies, such as high gas pressures and densities \citep[e.g.][]{Elmegreen_and_Efremov_1997}. This has led to the formulation of a family of models in which GCs are the byproduct of normal star and galaxy formation throughout cosmic history \citep[e.g.][]{Kruijssen_2015,Keller_etal_2020}.

In our current standard paradigm for structure formation, galaxies form at the
centre of dark matter haloes that collapse in a bottom-up fashion, with small
systems forming first and later merging into progressively more massive
structures. In this framework, galaxy formation is a complex physical process
that involves both gas condensation at the centre of dark matter haloes, as
well as galaxy mergers and interactions either with other galaxies or with the
central regions of dark matter haloes \citep*[for a classical reference, see e.g.][]{Mo_vandenBosch_White_book}. This means that an end-to-end
description of the formation process of today's GCs should include
an explicit treatment of both their formation and their dynamical evolution
within their evolving host galaxy haloes.

In the past years, different attempts have been made to study the dynamical
evolution of GCs within their host galaxy haloes. These include largely analytical studies that
focused on the effects of two-body relaxation, gravitational shocks and mass
loss by stellar evolution on the mass function of star clusters, starting from
an initial distribution approximated by a power-law
\citep[e.g.][]{Fall_and_Zhang_2001,Prieto_and_Gnedin_2008,Elmegreen_2010,Kruijssen_2015}; work that has
tried to constrain the physical processes leading to the formation of GCs by
using their observed metallicity distributions in the local Universe and a combination of empirical relations and/or merger trees extracted from N-body simulations  
\citep[e.g.][]{Tonini_2013,Li_and_Gnedin_2014,Choksi_etal_2018,Chen_and_Gnedin_2022}; and other empirical approaches that combine numerical merger trees with the assumption, based on observations in the local Universe, of a power-law relation between halo mass and mass in GCs \citep{Ramos-Almendares_etal_2020,Valenzuela_etal_2021}. This previous work either
focused on specific aspects of GC evolution/formation or neglected important
physical mechanisms of GC evolution.

Resolving the process of GCs formation directly within galaxy formation
simulations is prohibitively expensive, because it would require extremely high
resolution (particle masses/cells below $10^3\,{\rm M}_{\sun}$, and sub-parsec force resolution to resolve the bulk of the GC population and the scales at which they form), as well as an appropriate treatment of the star formation and stellar feedback processes. While some work has begun to resolve aspects of GC formation in
cosmological simulations
\citep[e.g.][]{Mandelker_etal_2017,Kim_etal_2018,Meng_and_Gnedin_2020,Ma_etal_2020,Lahen_etal_2020,Li_etal_2022},
the approach remains limited to small volumes and a narrow range in redshifts, preventing a
detailed comparison with the wealth of observational data in the local
Universe.

An alternative approach is that of modelling GC formation and evolution within
their parent galaxy haloes resorting to `sub-grid' or `semi-analytic'
models. The important advantage, in this case, is that the limited
computational costs allow an efficient investigation of the influence of
different specific assumptions, as well as a rapid exploration of the
parameter space. Coupling these techniques to dark matter-only and
high-resolution cosmological volumes provides access to a large dynamic range in
halo and galaxy masses allowing statistical analysis as a function of redshift,
galaxy properties, and environment. Efforts in this direction include 
post-processing analyses of dark matter simulations with the inclusion of
baryons through scaling relations inspired by observational data or physical models
\citep*{Tonini_2013,Kruijssen_2015,Choksi_etal_2018,El-Badry_etal_2019}.

In more recent years, direct simulations of galaxy formation have been used to
model the formation and evolution of GCs. For instance, the E-MOSAICS project
\citep{Pfeffer_etal_2018,Kruijssen_etal_2019} couples an analytic model that
describes the formation, evolution, and disruption of stellar clusters to the
EAGLE galaxy formation model. Over the past years, the E-MOSAICS simulations have been used to provide context, interpretation, and predictions for a wide range of GC properties, such as their numbers \citep[e.g.][]{Bastian_etal_2020}, metallicity distribution \citep[e.g.][]{Usher_etal_2018,Pfeffer_etal_2023}, formation histories \citep{Reina-Campos_etal_2019}, mass function \citep{Hughes_etal_2022}, spatial distribution and kinematics \citep{Trujillo-Gomez_etal_2021,Reina-Campos_etal_2022}, origin \citep{Pfeffer_etal_2019,Trujillo-Gomez_etal_2023}, and their use in tracing galaxy formation and assembly \citep{Hughes_etal_2019,Kruijssen_etal_2019,Kruijssen_etal_2020,Pfeffer_etal_2020,Dolfi_etal_2022}. The success of this approach has motivated several similar initiatives \citep[e.g.][]{Reina-Campos_etal_2022b,Rodriguez_etal_2022,Doppel_etal_2023,Grudic_etal_2023}. The big hurdle faced by all of these models is one of statistics, as the requirement of resolving galaxy formation implies that cosmological volumes larger than $\sim50$~Mpc \citep[see e.g.][]{Bastian_etal_2020} remain out of reach.

In this work, we adopt an approach very similar to that used in the E-MOSAICS
project, but take advantage of a state-of-the-art semi-analytic galaxy
formation model to describe the evolution of the galaxy
population across a much larger cosmological volume that can be spanned by spatially resolved hydrodynamical simulations. Specifically, we build on the GC model presented in
\citet{Kruijssen_2015} that explains the observed properties of GCs as the
natural outcome of star and cluster formation in high-redshift galaxies, and include
its basic assumptions in the GAlaxy Evolution and Assembly (GAEA) semi-analytic
model \citep{DeLucia_etal_2014,Hirschmann_etal_2016}, coupled to a large dark matter-only
cosmological simulation. In this paper, we provide the details of our model and
discuss how its basic predictions for the GC population compare with available
data.

The layout of the paper is as follows: we present the simulation and the galaxy
formation model used in our study in Section~\ref{sec:simsam}.
Section~\ref{sec:starcl} provides a detailed description of how we have
included in our semi-analytic model the formation of young stellar clusters, how we have
modelled their evolution, and how we have tested various physical descriptions of these physics. In Section~\ref{sec:ex}, we present a case study of two model galaxies to illustrate how the mass distribution of GCs evolves as a function of time. Sections~\ref{sec:basicrels} and \ref{sec:basicdistr} show the
basic predictions of our model and compare them to observational estimates.
Finally, in Section~\ref{sec:discconcl}, we discuss our results and present our
conclusions.

%%%%%%%%%%%%%%%%%%%%%%%%%%%%%%%%%%%%%%%%%%%%%%%%%%%%%%%%%%%%%%%%%%%%%%%%%%%%%%%

\section{The simulation and the galaxy formation model}
\label{sec:simsam}

The model predictions presented in this work are based on dark matter merger
trees extracted from the Millennium Simulation \citep{Springel_etal_2005}. This
dark matter-only simulation follows 2,160$^3$ particles in a comoving box of 500~${\rm
  Mpc}\,{\rm h}^{-1}$ on a side, and assumes cosmological parameters consistent
with WMAP1 ($\Omega_\Lambda=0.75$, $\Omega_{\rm m}=0.25$, $\Omega_b=0.045$, $n=1$,
$\sigma_8=0.9$, and $H_0=73 \, {\rm km\,s^{-1}\,Mpc^{-1}}$). In previous work
\citep{Wang_etal_2008}, we have shown that (small) modifications of the
cosmological parameters do not significantly affect model predictions, once the
model parameters are retuned to reproduce a given set of observational results
in the local Universe. Therefore, we do not expect significant changes with an
updated cosmological model. We will verify this in future work, where we plan to extend our analysis to
higher resolution simulations and an updated cosmological model.

In this work, we take advantage of the GAlaxy Evolution and Assembly ({\sc
  GAEA}) model of galaxy formation\footnote{https://sites.google.com/inaf.it/gaea/}, described in detail in
\citet{Hirschmann_etal_2016}, with the updated modelling for disk sizes described in \citet{Xie_etal_2017}. While the latter modification does not affect significantly the basic predictions of our model, it leads to a better agreement between model predictions and observational data for galaxy sizes - an element that we deem important for the stellar cluster model discussed in this paper. We refer to \citet{Hirschmann_etal_2016} and references therein for full details about the modelling adopted for all physical processes considered. Briefly, we have four different baryonic components (plus an equivalent number of metal components) associated to each model galaxies: hot gas at temperature $>10^4$~K that; cold gas that is only associated with galaxy disks in which star formation takes place; stars; and an ejected component that stores material that is outside the star forming phase because of stellar feedback. Star formation occurs, in galaxy disks, at a rate that is proportional to the amount of cold gas available above a surface density threshold that is modelled following \citet{DeLucia_and_Helmi_2008}. The interstellar medium of galaxies is enriched by metals released by stellar winds and supernovae explosions. Our galaxy formation model includes a sophisticated treatment for the non-instantaneous recycling of gas, metals and energy that accounts for the finite lifetimes of stars and allows us to follow individual metal abundances \citep[full details about the approach adopted can be found in][]{DeLucia_etal_2014}. The adopted stellar feedback model is partly based on scaling relations extracted from hydro-dynamical simulations. Specifically, these relations are used to parametrize the rate at which cold gas in disks can be reheated and then, assuming energy conservation arguments, to quantify the rate at which reheated gas can be ejected in a galactic outflow. This ejected component can be eventually re-incorporated onto the hot gas, on a time-scale that depends on the halo mass, and can then later cool onto the central galaxy of the halo. Full details about the stellar feedback scheme can be found in  \citet{Hirschmann_etal_2016}.  Galaxy mergers trigger starbursts episodes and lead to the formation of stellar bulges; the gas converted into stars is proportional to the baryonic mass ratio of the two merging galaxies, with model parameters being calibrated against results from controlled numerical experiments. We further assume that in case of major mergers, that occur when the mass ratio is larger than 0.3, the disc is completely disrupted and all the stars put in a bulge component. Full details about these prescriptions can be found in \citet{DeLucia_and_Blaizot_2007} and \citet{DeLucia_etal_2010}. 

In previous work, we have shown that
our reference model is able to reproduce a large number of important
observational constraints, including the galaxy stellar mass function up to
$z\sim 7$ and the cosmic star formation rate density up to $z\sim 10$
\citep{Fontanot_etal_2017}, the relation between galaxy stellar mass and gas
metallicity and its secondary dependence as a function of star formation rate
and gas mass \citep{DeLucia_etal_2020}, as well as the observed evolution of
the galaxy mass - gas/star metallicity relations
\citep{Hirschmann_etal_2016,Fontanot_etal_2021}.

For the analysis presented in this paper, we have used only a small fraction
(about 10~per~cent) of the entire volume of the Millennium Simulation. On the
basis of previous work, we consider this sub-volume representative (it includes
a few of the most massive haloes that can be identified in the entire
simulation volume). In addition, we have used the same parameter set adopted in
\citet{Hirschmann_etal_2016} with only one modification: in our published
reference model, we had assumed that galaxies at the centres of haloes with masses
below $5\times 10^{10}\,{\rm M}_{\sun}$ would inject most of the newly
synthesized metals (95~per~cent) directly into the hot gas phase. This
modification had been included in previous work focused on faint satellites of
our Milky-Way \citep{Li_Yang-Shyang_Helmi_2010} to better reproduce their
  metallicities, and was motivated by results of hydro-dynamical simulations of dwarf galaxies \citep{MacLow_and_Ferrara_1999}. While not affecting significantly galaxy (and GC) properties
on larger scales, it does enter a critical resolution regime for the dark
matter simulation used in this work: the particle mass is $m_{\rm
    p}=8.6\times10^8\,{\rm M}_{\odot}$, which means that a halo of
  $5\times10^{10}\,{\rm M}_{\odot}$ is resolved with only about 40
  particles. Therefore, in this work we assume that all new metals are
immediately mixed with the cold gas in the disc, and explicitly show when this
has some influence on our model predictions (see Section~\ref{sec:basicdistr}).

%%%%%%%%%%%%%%%%%%%%%%%%%%%%%%%%%%%%%%%%%%%%%%%%%%%%%%%%%%%%%%%%%%%%%%%%%%%%%%%
\section{A two-phase model for the origin of globular clusters}
\label{sec:starcl}

To model the abundance and properties of GCs, we build on the
`two-phase' model introduced by \citet{Kruijssen_2015}, including additional
implementations that we detail in the following. We refer to the original study
by \citet{Kruijssen_2015} for a detailed review of the approach, which attempts
to provide an end-to-end description of the origin of GCs, from their formation
at high redshift until the present day. In summary:
\begin{itemize}
\item[(i)] young stellar clusters are assumed to form in the high-pressure discs
  hosted by high-redshift galaxies;
\item[(ii)] the clusters undergo rapid disruption by tidal perturbations from
  molecular clouds and clumps in the host galaxy disc; 
\item[(iii)] the disruption continues until clusters migrate into the halo of
  the galaxy, following e.g.\ a merger event;
\item[(iv)] finally, clusters undergo a slow evolutionary phase, in the host
  galaxy halo, due to tidal evaporation. 
\end{itemize}

In the following subsections, we describe the specific prescriptions adopted to
include each of the four elements listed above in our galaxy formation
model. As we explain in detail below, we assume stellar clusters form in the host galaxy disk, and migrate into the galaxy halo after galaxy mergers. In practice, we have associated to each model galaxy two
three-dimensional arrays storing the information about the stellar clusters
that are associated with the disk and the halo of the galaxy. The three
dimensions of each array correspond to bins of stellar cluster mass (we
consider \texttt{NMBINS}$=$~50 mass bins, logarithmically spaced between $10^2$
and $10^{10}\,{\rm M}_{\odot}$), [Fe/H] (\texttt{NZBINS}$=$~20, linearly spaced
between -3.5 and 0.68), and formation time (\texttt{NTBINS}$=$~27, linearly spaced
between 0 and 13~Gyr). The disk and halo arrays are initialized and evolved as
detailed in the next sections.

\subsection{Formation of young stellar clusters}
\label{sec:clinit}

In our galaxy formation model, star formation can take place in a `quiescent'
mode (from cold gas associated with the galaxy disk), and during merger driven
starbursts. We assume that a (small) fraction of the star formation occurring only
through the quiescent channel leads to the formation of young massive clusters, that inherit the metallicity of the star forming gas. We suppress the formation of new stellar clusters during merger-driven starbursts, as they are expected to be associated with very strong tidal perturbations (See Section 3.3). A new population of stellar clusters is initialized for each new episode of
star formation: if $\Delta M_{\rm star}$ is the mass of stars formed during a code
time-step\footnote{The differential equations governing the evolution of the baryonic components associated with model galaxies are solved dividing the time interval between two subsequent
  snapshots (the available simulation outputs) into 20 time-steps, each corresponding to $\sim 10-19$~Myrs up to
  $z=3$, and even smaller timescales at higher redshift (where the snapshots are closer in time). We have verified, in previous work, that the number of sub-steps adopted is enough to achieve numerical convergence.}, we assume that the
corresponding mass in stellar clusters is $\Gamma \Delta {\rm M}_*$, where
$\Gamma$ quantifies an environmentally dependent cluster formation efficiency
(CFE). We have computed this quantity using the analytic model presented in
\citet{Kruijssen_2012}, in which bound stellar clusters collapse in the highest-density regions of the interstellar medium. In the framework of this model, the CFE can be expressed as the product of the fraction of star formation that results in bound structure and the fraction thereof that survives the `cruel cradle effect', i.e. the tidal disruption of star-forming regions or young stellar clusters by encounters with in the natal environment. We refer to the original paper by \citet{Kruijssen_2012} for full details on the derivation of these quantities. 

The CFE is determined by galaxy scale physics, and can be expressed in terms of
the cold gas surface density, the galaxy angular velocity, and the \citet{Toomre_1964} $Q$ instability parameter. For this calculation, and throughout this paper, we
have set the Toomre instability parameter ($Q$) equal to 1. This value is
within the typical range observed in star-forming galaxies at $z\sim 2$,
$Q=0.2-1.6$ \citep{Genzel_etal_2014}. The cold gas surface density and galaxy angular velocity are given by:
\begin{displaymath}
\Sigma_{\rm gas} = M_{\rm cold} / \pi r_{\rm half}^2 \,\,\,\,\,\, {\rm
  and} \,\,\,\,\,\, \Omega = V_{\rm max}/r_{\rm half}.
\end{displaymath}
In the above equations, $M_{\rm cold}$ is the cold gas mass, and $r_{\rm half}$
is the disk half mass radius. The latter is computed as $1.68\times R_{\rm d}$
(assuming an exponential disk), where $R_{\rm d}$ is the scale radius of the
disk and is computed by tracing the specific angular momentum of the gas as
described in detail in \citet{Xie_etal_2017}. $V_{\rm max}$ is inherited from
the subhalo catalogue and is the maximum circular velocity of the parent
dark matter substructure for each model galaxy (for orphan satellites,
i.e.\ those that are no longer associated with an existing substructure, this is
the value at the last time there was a distinct parent subhalo).
%, tracing a more fundamental dependence on the gas pressure. 
The CFE resulting from the model by \citet{Kruijssen_2012} is found to increase with the gas surface density, from $\Gamma\sim1$~per cent in low-density galaxies up to $\Gamma\sim 70$ for surface gas densities $\sim 10^3\,{\rm M_{\sun}\,{\rm pc}^{-2}}$. This provides a good match to cluster formation
efficiencies that are observed in the local Universe, ranging from a few per
cent to as much as 50~per~cent in high-density starburst environments
\citep[e.g.][]{Adamo_etal_2015,Adamo_etal_2020,Adamo_etal_2020b,Johnson_etal_2016}.

A proper calculation of $\Gamma$ requires several integrations that would slow
down significantly our galaxy evolution code. In order to speed up our
computation, we have used tabulated values of $\Gamma$ (on a $30\times30$ grid)
corresponding to different values of the gas surface density and of the angular
velocity. The entries of the table are then linearly interpolated at each star
formation episode, to compute the corresponding CFE. We have verified that
values obtained using our approach are very close to those that would be
obtained using a full integration calculation.

Observations of young stellar cluster populations \citep{Zhang_and_Fall_1999,Hunter_etal_2003,Larsen_2009,PortegiesZwart_etal_2010} have shown that the initial cluster mass function (ICMF) is well described by a power law with index $\alpha=-2$, which is consistent with expectations from gravitational collapse in hierarchically structured clouds \citep{Elmegreen_and_Falgarone_1996,Guszejnov_etal_2018}. At low masses, a fiducial truncation corresponding to $M_{\rm min}=100\,{\rm M}_{\sun}$ is typically adopted \citep[][and references therein]{Lada_and_Lada_2003,Kruijssen_2015}. However, this assumption appears in contrast with findings that very large fractions (up to about $50$ per cent) of low-metallicity stars in some nearby dwarf galaxies are bound to their GCs \citep{Larsen_etal_2018}, which implies that few low-mass clusters (or possibly none) could have formed coevally with these GCs. In fact, an ICMF extending down to $100\,{\rm M}_{\sun}$ in these galaxies would require the majority of low-mass clusters to have been disrupted, returning their mass to the field population. This would allow for a maximum of 10~per~cent of the low-metallicity stars to reside in surviving GCs, contrary to observations. At the high-mass end, an exponential truncation of the ICMF is often assumed. Observational studies find truncation masses varying from $M_{\rm max} = 0.5$ to $10^5\,{\rm M}_{\sun}$ \citep{Gieles_etal_2006,Larsen_2009,Bastian_etal_2012,Konstantopoulos_etal_2013}.

In this work, we assume that the ICMF is well-described by a power law, with
exponential truncations at both the high- and low-mass ends \citep{Trujillo-Gomez_etal_2019}:
\begin{equation}
\frac{{\rm d}N}{{\rm d}M} \propto M_i^{\alpha} {\rm exp}\left(
\frac{-M_{\rm min}}{M} \right) {\rm exp}\left( \frac{-M}{M_{\rm max}} \right). 
\label{eq:cimf}
\end{equation}
We assume $\alpha=-2$, and that $M_{\rm max}$ is determined by
evaluating the mass fraction of a centrifugally limited region that can
collapse before stellar feedback is able to halt star formation, as detailed in
\citet*{Reina-Campos_and_Kruijssen_2017}. In this study, the
resulting upper truncation mass is expressed in terms of the cold gas surface density, the galactic angular velocity, and the Toomre parameter $Q$, with the latter having a limited impact on the cluster mass scale.  The model predicts that more massive clusters are formed in environments characterized by larger gas surface densities, providing a natural explanation for the existence of more massive clumps at higher redshift.

As for $M_{\rm min}$, we adopt the model presented
in \citet*{Trujillo-Gomez_etal_2019}, based on empirical scaling relations of
molecular clouds in the local Universe, and on the hierarchical nature of star
formation in clouds regulated by stellar feedback.  Specifically, the minimum
mass of stellar clusters is evaluated by estimating the time-scale required for
stellar feedback to halt star formation, against the collapse time of clouds
with a given mass spectrum. This allows an estimate of the range of cloud
masses that can achieve the minimum star formation efficiency required to
remain bound after feedback has blown out the remaining gas. The resulting
values of $M_{\min}$ vary by orders of magnitude from local and quiescent discs
to high-redshift starbursting systems, with a strong dependence on the surface
density of the ISM and no significant dependence on the
galaxy angular velocity and the Toomre parameter $Q$. Below, we will test
explicitly the impact of assuming an environmentally dependent $M_{\min}$.

Both $M_{\min}$ and $M_{\max}$ can be expressed as a function of the gas
surface density and of the galaxy angular velocity, as in the case of CFE, which for all of these quantities reflects an underlying physical dependence on the gas pressure. For the
sake of computational time, we have generated pre-computed tables giving the
minimum and maximum truncation mass over the same grid used to pre-compute
$\Gamma$.  The tables are linearly interpolated to evaluate Eq.~\ref{eq:cimf}
each time the cluster mass function needs to be initialized or updated, due to
a new episode of star formation.
%% a truncation mass $M_c$, and a lower mass limit of
%% $M_{\rm min}= 10^2\,{\rm M_{\odot}}$ \citep[][and references
%% therein]{Lada_and_Lada_2003,Kruijssen_2015}. The truncation mass is
%% assumed to be proportional to the Toomre
%% mass \citep{Kruijssen_2014,Adamo_etal_2015}:
%% \begin{displaymath}
%% M_c = \epsilon \, \Gamma M_{\rm T}
%% \end{displaymath}
%% where $\epsilon = 0.03$ corresponds to the star formation efficiency
%% parameter
%% adopted in our model, and the \citet{Toomre_1964} mass is computed as:
%% \begin{displaymath}
%% M_{\rm T} = \pi^4 G^2 \Sigma^3_{\rm gas}/4\Omega^4.
%% \end{displaymath}
%% $\Sigma_{\rm gas}$ and $\Omega$ are the gas surface density and the angular
%% velocity of the galaxy, respectively. 

\subsection{Cluster evolution during the rapid disruption phase}
\label{sec:rd}

\begin{figure*}
\centering
\resizebox{16cm}{!}{\includegraphics{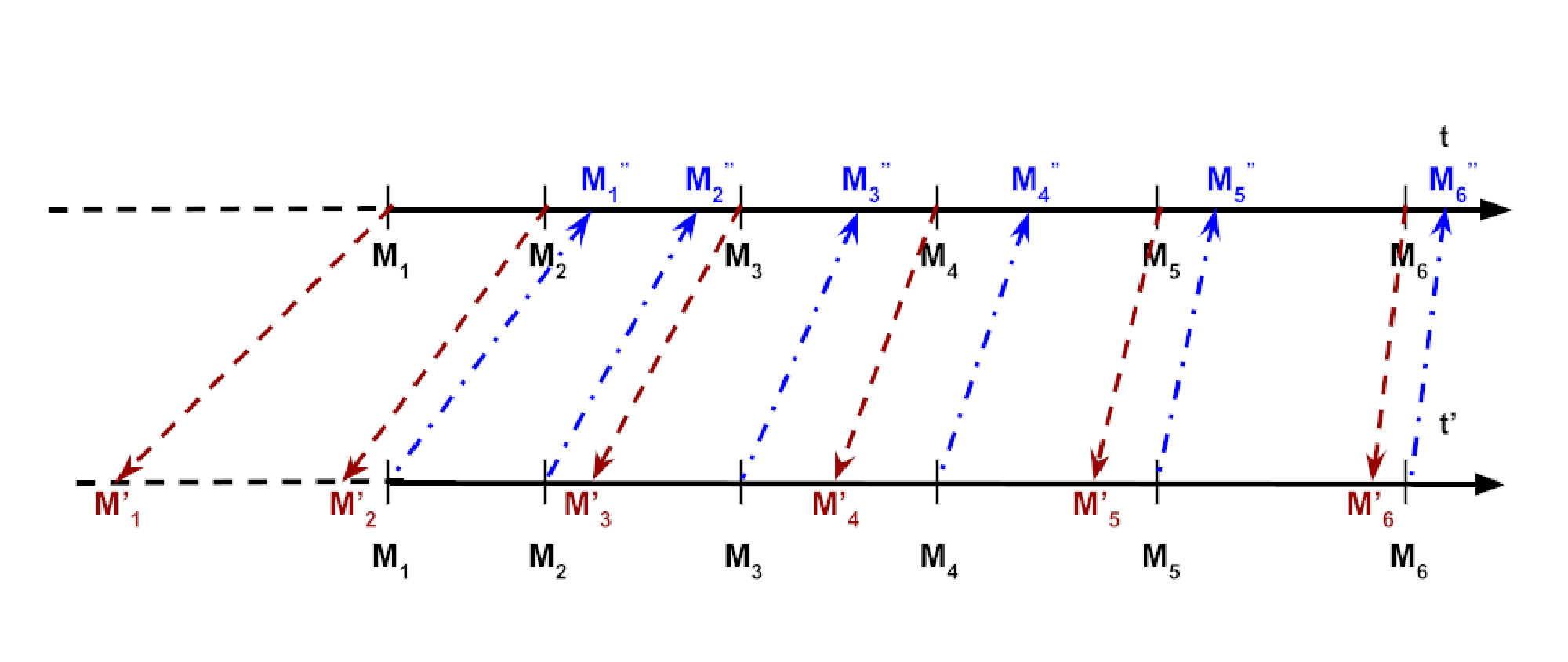}} 
\caption{A sketch of our implementation of the rapid disruption evolutionary
  phase of star clusters in our galaxy formation model. The two horizontal
  lines show the mass projection of the three-dimensional array associated with
  each model galaxy, at two subsequent time-steps. ${\rm M}_{i}$ (black, with
  $i=1,2,...6$) represent the boundaries of the mass bins considered;
  ${\rm M}'_{i}$ (red) show the corresponding evolved values at the following
  time-step $t'$ (these are computed using Eqs.~\ref{eq:dmtcce} and
  \ref{eq:tcce}). Finally, ${\rm M}''_{i}$ (blue) correspond to the mass values
  that would evolve into the fixed grid values at $t'$ (see text for
  details). The latter values can be used to compute the evolved cluster mass
  function at time $t'$ from the unevolved distribution at time $t$.
  \label{fig:rd}}
\end{figure*}

After their formation, stellar clusters evolve within the host galaxy disk, and
are subject to frequent and strong tidal perturbations by dense gas clumps \citep[e.g.][]{Gieles_etal_2006,Elmegreen_and_Hunter_2010,Kruijssen_etal_2011}.
Following \citet[][see original paper and references therein]{Kruijssen_2015}, we describe the rapid disruption phase driven by these tidal shocks in
terms of a mass loss rate:
\begin{equation}
({\rm d}M / {\rm d}t)_{\rm cce} = - M / t_{\rm cce},
\label{eq:dmtcce}
\end{equation}
where the subscript `cce' stands for `cruel cradle effect'
\citep{Kruijssen_etal_2012b}. The disruption
time-scale can be expressed as \citep{Kruijssen_2015}:
\begin{equation}
t_{\rm cce} = t_{5{\rm ,
cce}} \Big(\frac{f_{\Sigma}}{4}\Big)^{-1} \Big(\frac{\rho_{\rm ISM}}{{\rm
M}_{\odot}\,{\rm pc}^{-3}}\Big)^{-3/2} \Big(\frac{M}{10^5\,{\rm
M}_\odot}\Big) \, \Phi^{-1}_{\rm ad}
\label{eq:tcce}
\end{equation}
where $t_{5{\rm , cce}} = 176$~Myr is a proportionality constant, $f_{\Sigma}$
is the ratio between the surface density of giant molecular clouds and the mean
gas surface density in the galactic mid-plane, and $\rho_{\rm ISM}$ is equated
to the mean density in a galactic disc mid-plane for an equilibrium disc
\citep{Krumholz_and_McKee_2005}:
\begin{displaymath}
\rho_{\rm ISM} = 3 \Omega^2 / \pi G.
\end{displaymath}
Finally, %we asssume a cluster radius that is independent of
% mass ($r_{\rm h} = 1.5\,{\rm pc}$).
$\Phi_{\rm ad}$ is a correction factor that accounts for the absorption of
tidal energy by adiabatic expansion. Following \citet{Kruijssen_2015}, we
assume:
\begin{displaymath}
f_{\Sigma} = 3.92 \Big( \frac{10 - 8 f_{\rm mol}}{2}\Big)^{1/2}
\end{displaymath}
with
\begin{displaymath}
f_{\rm mol} = 1 / (1+0.025\, \Sigma_{\rm gas,2}^{-2}).
\end{displaymath} 
In the last equation, $\Sigma_{\rm gas,2}$ is the surface density of the gas in
units of $100\,{\rm M}_\odot\,{\rm pc}^{-2}$.
The adiabatic correction is expressed as follows:
\begin{displaymath}
\Phi_{\rm ad} = \Big[ 1 + 9 \Big( \frac{\rho_{\rm h}/\rho_{\rm
ISM}}{10^4}\Big)\Big]^{-3/2}, 
\end{displaymath}
where $\rho_{\rm h} = 3M/8\pi r_{\rm h}^3$.

In the following, we will consider two different assumptions for the cluster
radius. In one case, we will assume that it is independent of the cluster mass,
and equal to $r_{\rm h} = 1.5\,{\rm pc}$. Alternatively, and this will be the
fiducial assumption in our reference run, we will assume that $r_{\rm h}$ depends on the
cluster mass. Specifically, we adopt eq.~13 of \citet{Gieles_and_Renaud_2016}:
\begin{equation}
  r_{\rm h} \simeq 3.8\,{\rm pc} \left( \frac{\gamma_{\rm GMC}}{12.8\,{\rm Gyr}}
  \right) ^{2/9} \left( \frac{M}{10^4 {\rm M}_{\odot}} \right) ^{1/9},
\label{eq:envradfirst}
\end{equation}
where
\begin{equation}
  \gamma_{\rm GMC} \simeq 6.5\,{\rm Gyr} \left( \frac{\sigma}{10\,{\rm km}\,
    {\rm s}^{-1}} \right) \left( \frac{10 \,{\rm M}_{\odot}^2 \,{\rm pc}^{-5}}
        {f_{\Sigma}\cdot \Sigma_{\rm gas} \cdot \rho_{\rm ISM}} \right)
\end{equation}
and, assuming an equilibrium disk and $Q=1$:
\begin{equation}
  \sigma = \frac{\pi}{2}\frac{G\,\Sigma_{\rm gas}}{\Omega}
\end{equation}
When considering environmentally dependent cluster radii, we multiply our expression
for $t_{\rm cce}$ by a factor:
\begin{equation}
(r_{\rm h}/r_{\rm h, 0})^{-3}, \,{\rm with }\, r_{\rm h, 0} = 1.5\,{\rm pc}.
\label{eq:envradlast}
\end{equation}
This makes disruption much faster (up to a factor 15) in low-density
environments, and is justified by the fact that more extended clusters would be
more susceptible to tidal perturbations, as observed.

To implement the rapid disruption phase in our galaxy formation model, we need
to account for the dependence of Eq.~\ref{eq:tcce} on both the stellar cluster
mass and on the physical properties of the host galaxy. Practically,
Eq.~\ref{eq:tcce} must be evaluated for each galaxy, at each time-step of the
evolution, and for each value of the cluster mass considered. To limit the
computational overhead, we have adopted an approach that is sketched in
Figure~\ref{fig:rd}, and that can be summarized as follows:
\begin{itemize}
\item[(i)] for each bin boundary of the mass grid considered at a given
  time $t$, we use Eqs.~\ref{eq:dmtcce} and \ref{eq:tcce} to compute the
  evolved values at the following time-step $t'$. In Figure~\ref{fig:rd}, this
  forward integration is illustrated by the dashed red lines, and ${\rm
    M}'_{i}$ (with $i=1,2,...6$) represent the evolved values of the grid 
    boundaries (${\rm M}_{i}$) at the time $t'$.
\item[(ii)] We then get the interpolated indices for which ${\rm M}'_i = {\rm
  M}_i$ at the time $t'$. These indices are used to compute the mass values at
  the time $t$ that would evolve into the fixed boundaries of the bins at
  the time $t'$. This `backward' integration is illustrated in
  Figure~\ref{fig:rd} by the dashed-dotted blue lines. ${\rm M}''_{i}$ (with
  $i=1,2,...6$) correspond to the values that would evolve into the fixed grid
  boundaries at the time $t'$.
\item[(iii)] The values obtained as described above can be used to compute the
  evolved cluster mass function at time $t'$ (i.e.\ the number of stellar
  clusters in each mass bin considered) starting from the mass distribution at
  time $t$. In the example shown in Figure~\ref{fig:rd}, the evolved number of
  stellar clusters in the first mass bin (between ${\rm M}_1$ and ${\rm M}_2$
  at time $t'$) can be computed by taking the number of stellar clusters in the
  unevolved array between ${\rm M}''_1$ and ${\rm M}''_2$ at time $t$. The
  evolved number of clusters in the second mass bin would be equal to the
  number of stellar clusters between ${\rm M}''_2$ and ${\rm M}''_3$ at the
  time $t$, etc.
\end{itemize}

To simplify our calculation, we assume that stellar clusters are distributed
uniformly\footnote{This assumption is correct in the case of infinitesimally small
  mass bins.} in each mass bin. We have verified that our results do not change
significantly when increasing the number of mass bins considered, so a more
sophisticated treatment is not expected to change significantly our results.

\subsection{Cluster migration into the galaxy halo}
\label{sec:merg}

The rapid disruption phase continues until the stellar clusters that form in
(and are associated with) the galaxy disk migrate into the halo. Following
\citet{Kruijssen_2015}, we assume that the migration agent is represented by
(minor and major) galaxy mergers. Major mergers occur when the baryonic mass
ratio between the merging galaxies is larger than 0.3. In this case, we assume
that all stellar clusters in the discs of both the accreted and accreting galaxies
migrate into the halo of the merger remnant. In case of a minor merger, we assume
that only the clusters in the disc of the accreted galaxy migrate into the halo
of the remnant, whereas the main progenitor instead retains the stellar clusters in its own disc.
During both minor and major mergers, the clusters in the haloes of both progenitor
galaxies are always transferred to the halo of the remnant galaxy.

In our model, galaxy mergers can trigger starbursts \citep[for details,
  see][]{DeLucia_and_Blaizot_2007,DeLucia_etal_2010}. We have suppressed the formation of a new
stellar cluster population during this star formation channel, because it is
expected to be associated with tidal perturbations stronger than those that
characterize quiescent star formation episodes. We note that this specific
assumption does not affect significantly our model results, as merger driven
starbursts only contribute to a minor fraction of the total star formation in
our model \citep{Wang_etal_2019}.

In addition, we also consider the possibility that the rapidly changing tidal
field during galaxy interactions can lead to an efficient cluster disruption in
the galaxy disc, before the merger is completed. To model this, we use results
from \citet{Kruijssen_etal_2012}, based on numerical simulations of merging
disc galaxies. In particular, we use their Eq.~10 to compute the survival
fraction of clusters (this is applied to all star clusters independently of their mass):
\begin{equation}
  f_{\rm surv} = 4.5\times 10^{-8}\,M_{\rm min,2}^2 \left( \frac{t_{\rm
      depl}}{\rm yr} \right)^{0.77 - 0.22{\rm log}(M_{\rm min,2})},
\label{eq:effdisr}
\end{equation}
and assume $M_{\rm min,2} = M_{\rm min}/10^2\,{\rm M}_{\odot}\sim 1$.  In the
work by \citet{Kruijssen_etal_2012}, $t_{\rm depl}$ is parametrized as the
ratio between the amount of gas available during the merger and the peak star
formation rate. We evaluate the latter quantity simply as the ratio between the
amount of stars formed during the burst associated with the merger and the corresponding internal code time-step.

\subsection{Cluster evolution during the slow disruption phase}
\label{sec:se}

After their migration into the galaxy halo, stellar clusters lose mass due to different physical processes: mass loss by stellar evolution and by dynamical effects, like two-body relaxation and stripping of stars due to the tidal field in which the star cluster is immersed and shocks \citep*[see e.g.][]{Lamers_etal_2010}. Following \citet[][we refer to the original paper for more details]{Kruijssen_2015}, we also describe the slow tidal evaporation phase in
terms of a mass loss rate:
\begin{equation}
({\rm d}M / {\rm d}t)_{\rm evap} = - M / t_{\rm evap},
\label{eq:dmtevap}
\end{equation}
with \citep{Lamers_etal_2005}:
\begin{equation}
t_{\rm evap} = t_{5{\rm , evap}} \Big( \frac{M}{10^5\,{\rm
M}_{\odot}}\Big)^{\gamma} 
\label{eq:tevap}
\end{equation}
In the last equation, $\gamma = 0.7$ and $t_{5{\rm , evap}}$ depends on the
stellar cluster metallicity:
\begin{displaymath}
{\rm [Fe/H]} = -1.03 - 0.5 \, {\rm log}(t_{5{\rm , evap}}/10\,{\rm Gyr}).
\end{displaymath}
This semi-empirical relation was adopted by \citet{Kruijssen_2015} so that the near-univeral characteristic mass-scale of GC is reproduced as a function of [Fe/H] at $z=0$. It reflects
the idea that the globular cluster metallicity correlates with the binding
energy of the galaxy in which it formed, which was proposed to be approximately preserved
during migration. The above relation between [Fe/H] and the disruption time is shown by \citet{Kruijssen_2015} to be consistent with the observed metallicity gradient of the Galactic GC population.

To implement the slow evaporation phase in our galaxy evolution model, we have
adopted an approach similar to that outlined in Section~\ref{sec:rd} for the
rapid disruption phase, but taking advantage of the fact that
Eq.~\ref{eq:tevap} does not depend on the physical properties of the host
galaxy and therefore does not need to be evaluated at each code
  time-step. From the practical point of view, this allows us to reduce the
number of calculations and pre-compute a mass loss grid at each new snapshot\footnote{This is
  necessary because the spacing between subsequent snapshots, and therefore
  also the internal time-step, are not constant. Otherwise, it would have been
  sufficient to compute the mapping only once.}
rather than for each internal time-step of our model. Specifically:
\begin{itemize}
\item for each boundary of the mass bins considered, and for each
  metallicity corresponding to central values of the metallicity bins
  considered, we have pre-computed the time-scales given by Eq.~\ref{eq:tevap}
  and the mapping needed to evolve the star clusters;
\item we have then used this pre-computed mass loss grid to evolve the mass
  function of the clusters associated with the galaxy halo at each internal
  time-step.
\end{itemize}

As for the rapid disruption phase, we assume that clusters are distributed
uniformly within each mass, metallicity, and formation time-bin. As mentioned
earlier, we have checked that results do not change significantly when
increasing the number of mass bins considered.

\subsection{Dynamical friction}
\label{sec:dftreat}

When computing the evolution of the most massive clusters in the galaxy disk, it is important to
account for the effect of dynamical friction, which can cause them to spiral in to the centre of the host galaxy. Following \citet[][see their
  Eq.~18]{Kruijssen_2015}, we evaluate the dynamical friction time-scale for
each central value of the mass bins considered:
\begin{displaymath}
  t_{\rm df} = 2\,{\rm Gyr} \,\Big(\frac{10^6 {\rm M}_{\sun}}{M} \Big) \,
  \Big(\frac{R}{2\,{\rm kpc}} \Big)^2 \,
  \Big(\frac{V}{200\,{\rm km}\,{\rm s}^{-1}} \Big)
\end{displaymath}
where we assume R is equal to the half-mass radius of the galaxy disk and
$V=V_{\rm max}$. We then simply set equal to zero the number of stellar
clusters in all those bins of our three-dimensional array that have experienced
more than one dynamical friction time-scale, integrated over their lifetimes.

As we will see in the following, our implementation of dynamical friction might
be too aggressive. This might be due to the fact that our implementation
  does not account for the mass loss of the clusters as they spiral in: the
  environment gets increasingly disruptive, which causes them to lose more mass
  and spiral in more slowly.

%%%%%%%%%%%%%%%%%%%%%%%%%%%%%%%%%%%%%%%%%%%%%%%%%%%%%%%%%%%%%%%%%%%%%%%%%%%%%%%
\subsection{Alternative physics}
\label{sec:altmodels}

To study the impact of the physics discussed in this section, we have
run alternative models in which we have switched off specific
processes one by one. In particular, we have considered the following cases:
\begin{itemize}
\item no environmental dependence of GC radii
  (Eqs.~\ref{eq:envradfirst}-\ref{eq:envradlast});
\item no efficient cluster disruption in the galaxy disc during galaxy mergers
  (see eq.~\ref{eq:effdisr});
\item no dynamical friction (see Section~\ref{sec:dftreat});
\end{itemize}
Additionally, we have considered a run where we assume that the minimum cluster mass
scale ($M_{\rm min}$) is always equal to $10^2\,{\rm M_{\odot}}$. We find that,
for the range of masses considered in this work, this assumption gives results
that are indistinguishable from those obtained using our reference run, that
assumes the model presented in \citet{Trujillo-Gomez_etal_2019} to determine
$M_{\rm min}$. Finally, as anticipated above, we have also considered a run
where metal ejection in small haloes is treated as in the model published in
\citet{Hirschmann_etal_2016}, i.e.\ almost all newly synthetized metals (95 per
cent) are directly injected into the hot gas component in haloes less massive
than $5\times 10^{10}\,{\rm M}_{\sun}$.

%%%%%%%%%%%%%%%%%%%%%%%%%%%%%%%%%%%%%%%%%%%%%%%%%%%%%%%%%%%%%%%%%%%%%%%%%%%%%%%
\section{Case studies: evolution of the GC mass function}
\label{sec:ex}

Before analysing the predictions of our model in detail, we discuss two `case
studies' to show how the mass distribution of stellar clusters evolves due to
the physical mechanisms discussed in Section~\ref{sec:starcl} and implemented
in our model. Our case studies correspond both to relatively low-mass galaxies because this allows us to visualize in some detail the evolution of their stellar cluster population. The corresponding history for more massive galaxies becomes more difficult to visualize, depending on the number of merger events suffered.

\begin{figure*}
\centering
\resizebox{16cm}{!}{\includegraphics{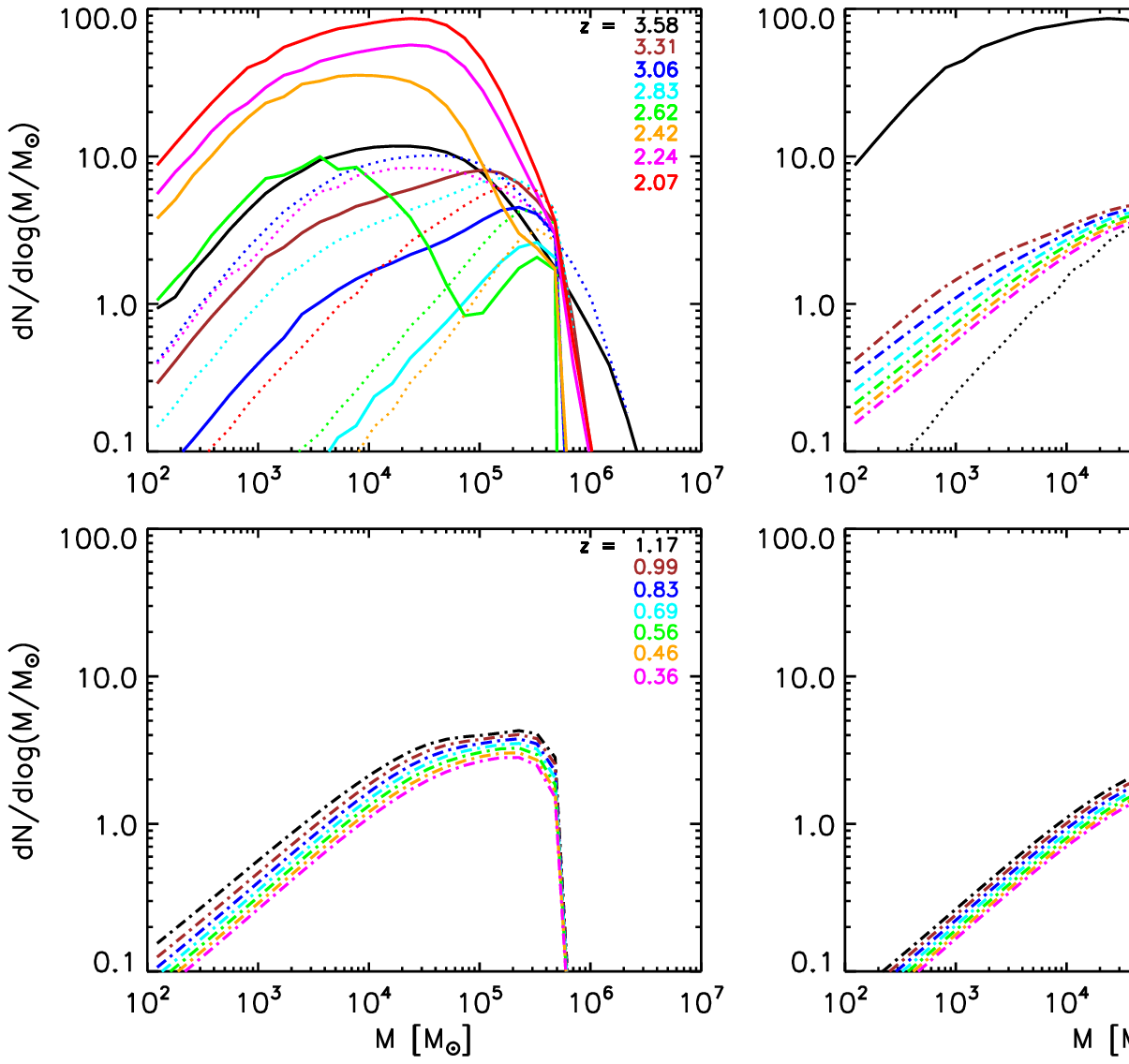}} 
\caption{Evolution of the mass distribution of stellar clusters associated with a model galaxy that has a stellar mass of $\sim 1.3\times10^9\,{\rm M}_{\sun}$ at $z=0$, and that experiences a single major merger at $z\sim 1.8$. The top left panel shows the mass distribution of stellar clusters associated with the two progenitors of the final galaxy, over the redshift range indicated in the legent. Solid and dotted lines indicate the stellar clusters associated with the disc of each progenitor, respectively. The top right panel shows again the mass distribution of the stellar clusters in the two progenitors right before the merger, and the evolution of the mass distribution of clusters associated with the halo of the remnant galaxy as dot-dashed lines. The same distribution, evolved down to $z=0$, is shown in the bottom panels.
\label{fig:ex1}}
\end{figure*}

Figure~\ref{fig:ex1} corresponds to a low-mass galaxy (the galaxy stellar mass at
$z=0$ is $\sim 1.3\times10^9\,{\rm M}_{\sun}$) that experiences one single
major merger event between $z=1.91$ and $z=2.07$ . The top left panel of Figure~\ref{fig:ex1}
shows the mass distribution of stellar clusters associated with the two
progenitors of the final galaxy at high redshift (a second progenitor exists only for those snapshots where a dotted line is plotted, i.e.\ from $z=3.06$ to
$z=2.07$). At these early times, the distribution of young stellar clusters
(those that are associated with the disk of our model galaxies) results from a
`competition' between ongoing star formation and the efficient rapid disruption of clusters by tidal shocks. At the final redshift shown in this panel, the total mass in stellar clusters is actually larger than that predicted at even higher redshift for the most massive progenitor (compare the red solid line with the black solid
line). The distributions shown in this panel also highlight the strong effect of our dynamical friction implementation, which we will return to below. The top-right panel again shows the distribution of stellar clusters in
both progenitors at $z=2.07$ (solid and dotted black lines) and the
distribution of clusters associated with the halo of the remnant galaxy after
the major merger (dot-dashed lines). As detailed in Section~\ref{sec:merg}, we
assume that the rapidly changing tidal field during the interaction leads to an
efficient cluster disruption before the stellar clusters associated to the
merging galaxies migrate to the halo of the remnant galaxy (dot-dashed
brown line). At this point, the evolution of stellar clusters is driven by
the slow evaporation phase described in Section~\ref{sec:se}. This slow
evolution of the cluster mass function can be appreciated, down to $z=0$, in
the bottom panels of Figure~\ref{fig:ex1}. At present, the mass distribution of
the clusters associated with the model galaxy considered in this example peaks
at $\sim 1.5\times10^5\,{\rm M}_{\sun}$, and corresponds to a specific frequency
of $\sim 13.8$ ($\sim 6.3$ when considering only stellar clusters more massive
than $10^5\,{M}_{\sun}$).

\begin{figure*}
\centering
\resizebox{16cm}{!}{\includegraphics{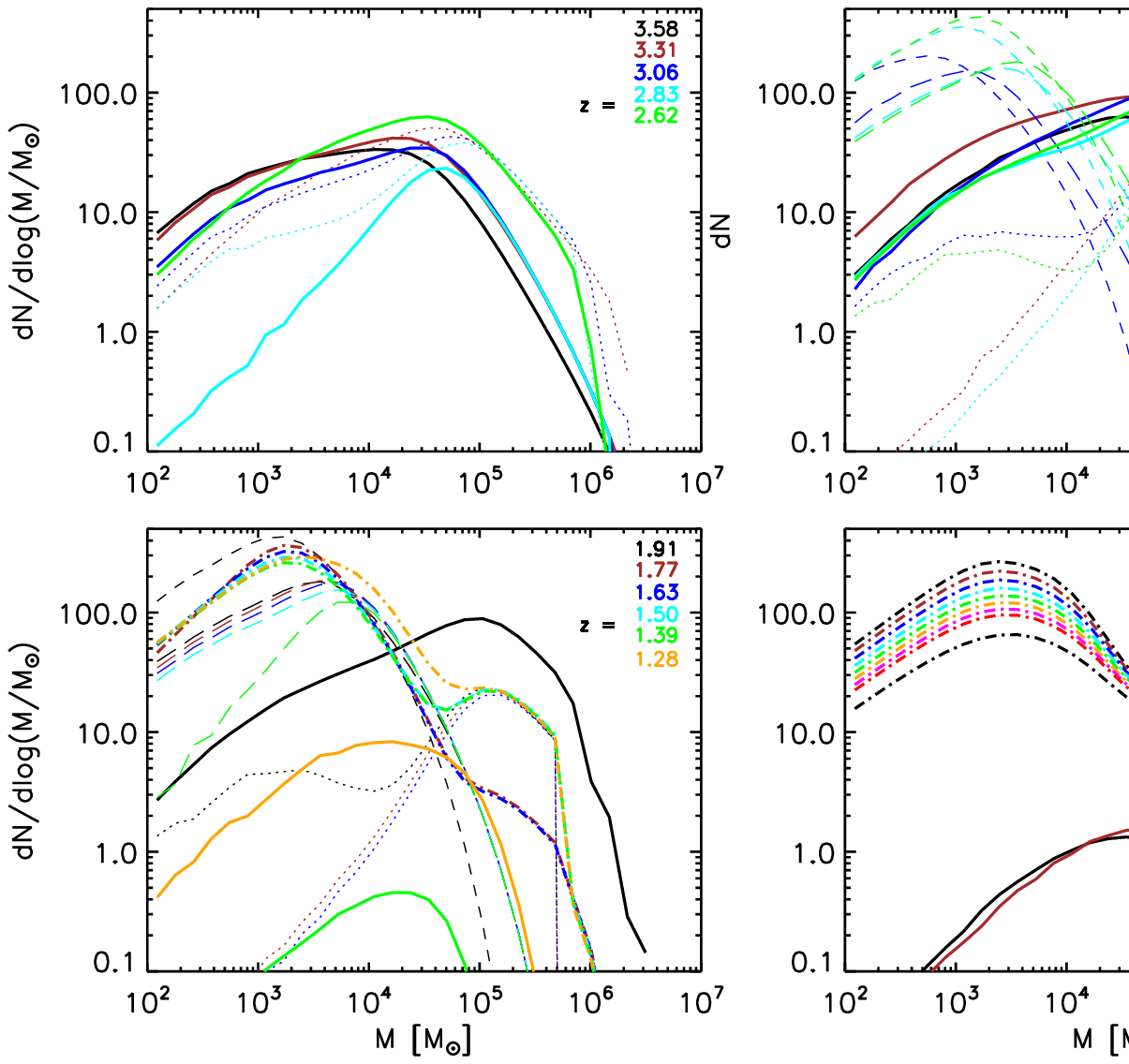}} 
\caption{As in Figure~\ref{fig:ex1}, but for a galaxy that has a stellar mass of $\sim 3.2\times10^9\,M_{\sun}$ at present, and that suffered one minor merger at $z\sim 1.6$ (with a galaxy of mass $\sim 2.7\times10^8\,{\rm M}_{\sun}$) and two major merger events at $z\sim 1.3$ $\sim 1.8$ (in this case, the merging galaxies have mass $\sim 3$ and $\sim 5\times10^8\,{\rm M}_{\sun}$, respectively). The mass distributions of stellar clusters disks are shown by solid, dotted, dashed and long dashed lines (solid is for the main progenitor, while other linestyles are used for other progenitors when present). The mass distribution of clusters in the halo is shown with dot-dashed lines.
\label{fig:ex2}}
\end{figure*}

Figure~\ref{fig:ex2} corresponds to a just slightly more massive galaxy (the
stellar mass at present is $\sim 3.2\times10^9\,M_{\sun}$) with a more
complex merger history. Specifically, the galaxy considered in this case has
suffered one minor merger episode at $z\sim1.5$ and two major mergers (one at
$z\sim1.8$ and the second at $z\sim1.3$). At the redshifts shown in the top
left and top right panels, there are two and four progenitors for this
galaxy. The mass distribution of star clusters in the disk evolves rapidly because of disruption and new star formation. The three merger events occurring during
the redshift interval shown in the bottom-left panel lead to the formation of a
stellar cluster component, associated with the halo of the model galaxy, which
then evolves slowly as a result of evaporation. At present, the mass distribution of
GCs associated with the example galaxy considered shows a double peak, one at a
mass slightly larger than $\sim 10^5\,{\rm M}_{\sun}$ and one at a mass $\sim
3.6\times 10^3\,{\rm M}_{\sun}$. The specific frequency of GCs measured for
this galaxy is $\sim 235$ if we consider all stellar clusters down to
$10^2\,{\rm M}_{\sun}$, and $\sim 16.5$ when we consider only stellar clusters
more massive than $10^5\,{M}_{\sun}$.

\begin{figure}
\centering
\resizebox{8cm}{!}{\includegraphics{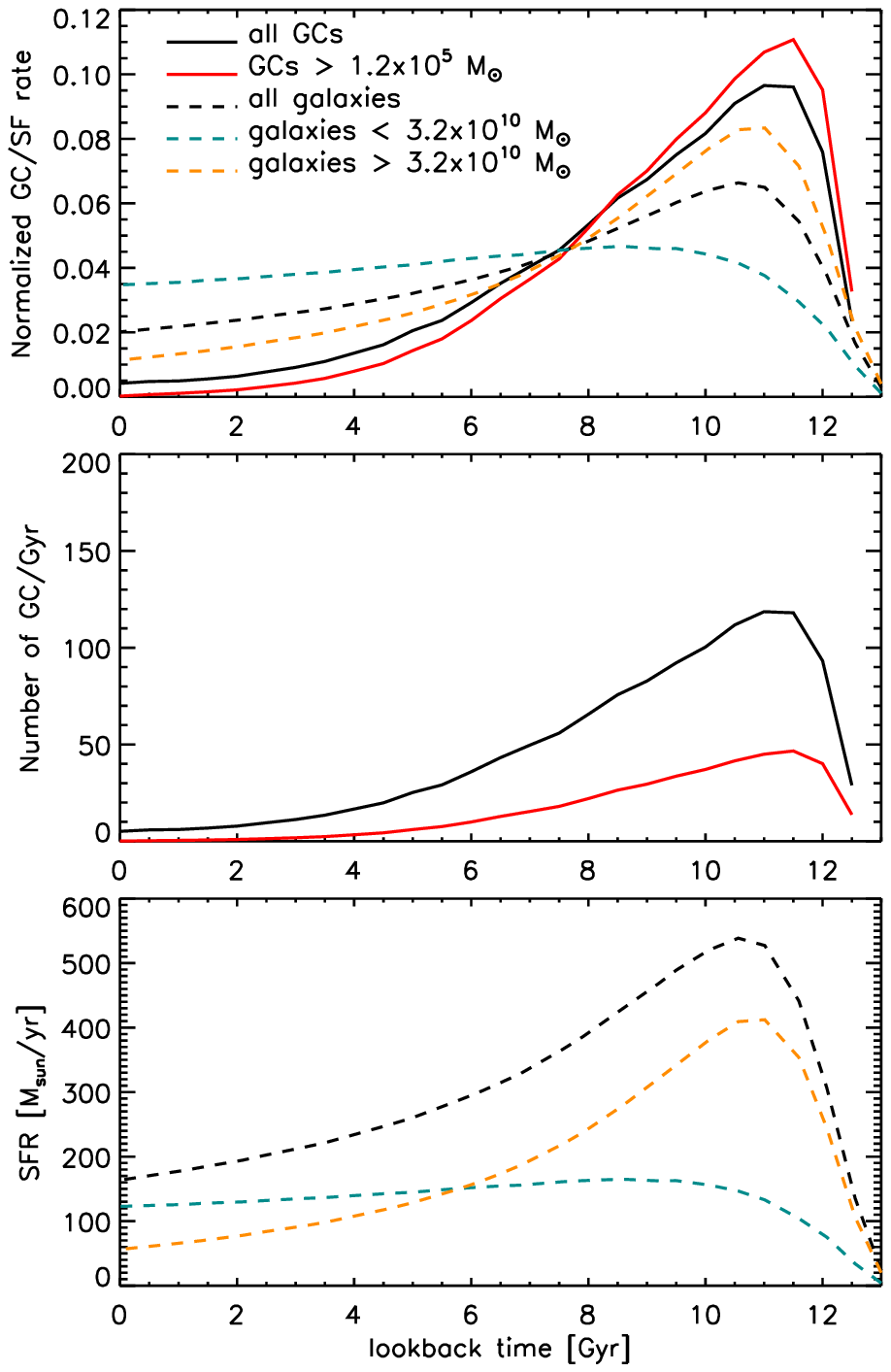}} 
\caption{Formation history of GCs (solid lines) compared to the star formation
  history of all model galaxies in the simulated volume considered. The dashed
  black lines correspond to galaxies more massive than $\sim 10^9\,{\rm
    M}_{\sun}$, while the orange and cyan lines correspond to galaxies more and
  less massive than $3.2\times 10^{10}\,{\rm M}_{\sun}$, respectively. In the
  top panel, all lines have been normalized to unity to ease their relative
  comparison, whereas the middle and bottom panels show the actual formation rate
  of GCs and stars, respectively.
\label{fig:sfh}}
\end{figure}

In the two example cases considered above, GCs form at early cosmic times and
migrate to the halo, where they are no longer subject to rapid disruption, at
$z>1-1.5$. Since the merger activity peaks at high redshift, most of the
surviving GCs associated with model galaxies will be old. Figure~\ref{fig:sfh}
compares the formation history of surviving GCs (solid lines, with the red line showing the formation history of GCs more massive than $10^5\,{M}_{\sun}$)
with the star formation history of all model galaxies in the simulated volume
considered. The black dashed line shows the average star formation history
obtained considering all model galaxies more massive than $10^9\,{\rm M}_{\sun}$, while the orange and cyan lines correspond to galaxies more and
less massive than $3.2\times 10^{10}\,{\rm M}_{\sun}$ respectively. All lines
have been normalised to unity in the top panel to ease the comparison, while the
middle and bottom panels show the actual values of the GC and star formation
rate, respectively. The figure shows that surviving GCs correspond to the oldest stellar component formed in our model galaxies: the formation history of GCs peaks at lookback time $\sim 11.5$~Gyr, about one Gyr earlier than the peak of the star formation when considering all surviving galaxies (above our mass limit). The figure also shows a trend for more massive GCs forming earlier than their less massive
counterparts. This is a survivor bias -- low-mass clusters are disrupted more rapidly, and any surviving low-mass GCs are therefore more likely to be young. In Section~\ref{sec:basicdistr}, we will analyse in more detail the
age (and metallicity) distribution of GCs associated with galaxies of different
stellar mass.

%%%%%%%%%%%%%%%%%%%%%%%%%%%%%%%%%%%%%%%%%%%%%%%%%%%%%%%%%%%%%%%%%%%%%%%%%%%%%%%
\section{Scaling relations and specific frequency of globular clusters}
\label{sec:basicrels}

Several observational studies have highlighted the existence of well-defined scaling relations between the GC population and the properties of their
host galaxies. In particular, the total mass in GCs scales almost linearly with
the host galaxy's halo mass across at least three or even four orders of
magnitude in halo mass \citep[from $\sim 10^{11}$ to $\sim 10^{15}\,{\rm
    M}_{\sun}$,][]{Blakeslee_etal_1997,Spitler_and_Forbes_2009,Hudson_etal_2014,Harris_etal_2017}. The
relation possibly extends for another two orders of magnitude (down to halo
masses $\sim 10^9\,{\rm M}_{\sun}$), with no significant deviation from
linearity but an increased scatter towards low halo masses
\citep{Forbes_etal_2018}.

The observed correlation appears surprising because, if GC formation is related
to star formation, one would expect a more natural correlation between the
total mass/number of GCs and the stellar mass of the galaxy. Virtually all studies
mentioned above interpreted the strong quasi-linear relation observed as an
indication that GCs formed at early times, with the process being unaffected by
the stellar/AGN feedback that introduces a non-linear correlation between
galaxy stellar mass and halo mass.  An alternative explanation was provided by
\citet{Kruijssen_2015}, who argued that the nearly constant ratio ($\eta$)
between the total mass in GCs and the halo mass arises from the combination of
galaxy formation within dark matter haloes and strongly
environmentally dependent cluster disruption. Recent work by
\citet{El-Badry_etal_2019}, based on a semi-analytic model for GC formation
coupled to dark matter merger trees, has argued that a constant value of $\eta$
is a natural consequence of hierarchical assembly, with the relation being
sensitive to the details of GC formation at low halo masses ($< 10^{11.5}\,{\rm
  M}_{\sun}$). These arguments have been revisited in a recent work by
\citet{Bastian_etal_2020}, based on the E-MOSAICS simulation suite. In
particular, the latter study argues that the normalisation of the relation is
primarily set by cluster disruption, while the downturn also predicted by
\citet{El-Badry_etal_2019} at low halo masses is imprinted by the underlying
relation between galaxy stellar mass and halo mass. 

\begin{figure*}
\centering
\resizebox{16cm}{!}{\includegraphics{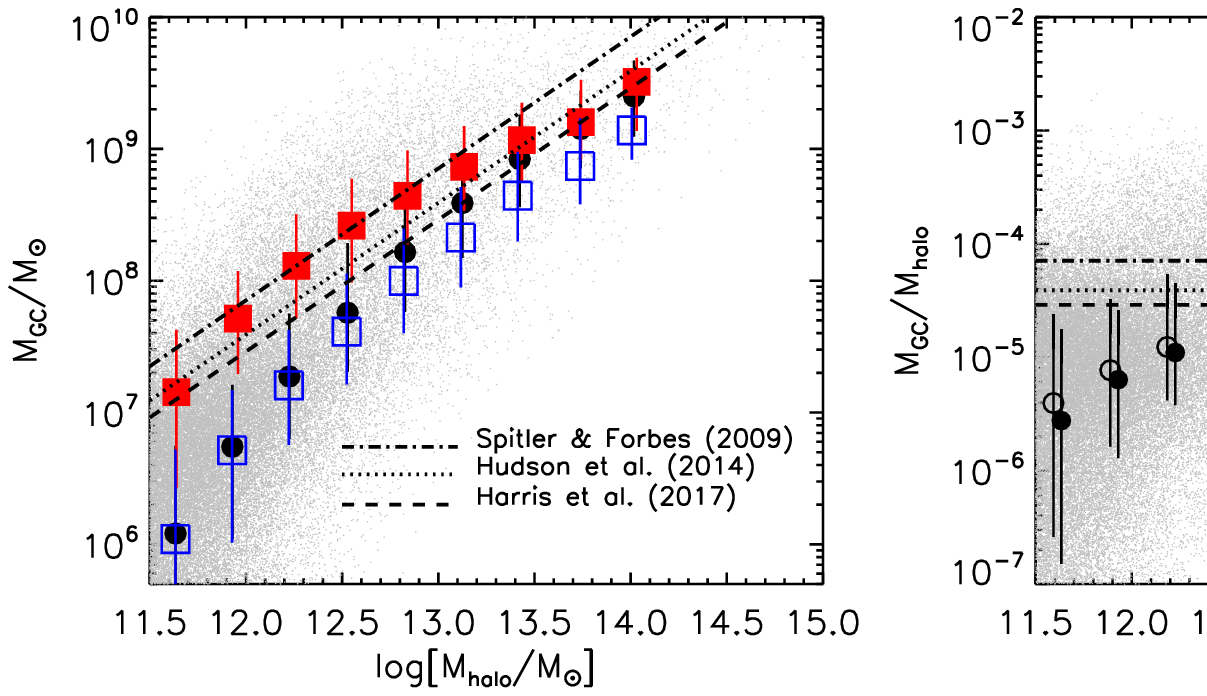}} 
\caption{Left panel: total mass in globular clusters as a function of the
  parent halo mass. Black symbols with error bars show the median and
  percentiles (25th and 75th) of the distributions obtained considering all
  central model galaxies. Individual values are shown as grey symbols. Red
  filled and blue open squares show the median and percentiles obtained for
  galaxies with bulge-to-total stellar mass ratio larger than 0.8 and smaller
  than 0.4, respectively. Different observational estimates are shown with
  lines of different styles, as indicated in the legend. Right panel: as for
  the left panel but this time showing the ratio between the total mass in
  globular clusters and halo mass. Empty symbols with error bars show the
  median and percentiles of the distribution obtained when including satellite
  galaxies.
\label{fig:MgcMhalo}}
\end{figure*}

Figure~\ref{fig:MgcMhalo} shows our model predictions. The left panel shows the
predicted correlation between the total mass in GCs and the parent halo mass,
for all model central galaxies in the simulated volume considered (solid
circles with error bars show the 25th and 75th percentiles of the
distributions). We have considered only central galaxies because observational
work typically uses scaling relations that are calibrated on central galaxies to
infer the parent halo mass. In addition, to have a fair comparison with
observational data, we considered all stellar clusters 
with masses $> 1.2\times10^5\,{\rm M}_{\sun}$, age $> 7$~Gyr, and
$-2.54<$~[Fe/H]~$<0.34$. These limits are motivated by the properties of the Galactic GC population \citep[e.g.][]{Kruijssen_etal_2019b}. We also show results
obtained when splitting our model galaxies into two subsamples according to
their bulge-to-total stellar mass ratio (solid and open squares correspond to
bulge/disc-dominated systems, respectively). In the right panel of the same
figure, we show how the ratio between the total mass in GCs and the halo mass
varies as a function of the halo mass. Empty symbols with error bars correspond to
the median and percentiles of the distribution obtained including also
satellite galaxies.  In this case, the plotted halo mass is obtained by
multiplying the number of bound particles associated with the parent dark
matter subhalo\footnote{For `orphan' galaxies, i.e.\ galaxies that are no longer
  associated with a distinct dark matter substructure, we have considered the
  number of particles associated with the last identified parent substructure.}
by the particle mass. Lines with different styles in both panels show observational estimates, as indicated in the legend.

Figure~\ref{fig:MgcMhalo} shows that our model predicts a quasi-linear relation
between the total mass in GCs and halo mass, for haloes more massive than $\sim
3\times 10^{12}\,{\rm M}_{\sun}$. A deviation from linearity is predicted for
lower-mass haloes. As mentioned above, a curvature at low halo masses has been found in previous work
\citep[e.g.][]{Choksi_etal_2018,El-Badry_etal_2019,Bastian_etal_2020}. In our model, the curvature is not affected by the alternative physical models considered (see Section~\ref{sec:altmodels} and the discussion below). The curvature is also only slightly weakened by the inclusion of satellite model galaxies. The latter has a more pronounced effect in the E-MOSAICS simulation suite \citep{Bastian_etal_2020}. We note here that haloes with mass $\sim 10^{12}\,{\rm M}_{\sun}$ are resolved with about $850$ particles in the Millennium Simulation, so it would be interesting to verify the trends predicted with a higher resolution volume -- a higher resolution would allow smaller mergers to be resolved, and these could potentially bring into the halo a larger number of GCs from lower-mass, accreted galaxies. In addition, one should bear in mind that observed halo masses of
low-mass galaxies are affected by relatively large uncertainties, which
limits the statistical significance of a downturn in observational data (see
discussion in \citealt{Bastian_etal_2020}).

Our model also predicts that bulge-dominated galaxies are typically characterised by a larger total mass in GCs compared to galaxies of similar
stellar mass but later morphological type. This is not surprising given
that the main channel for bulge formation in our model is represented by
galaxy mergers \citep{DeLucia_etal_2011}, and that we have assumed that mergers
trigger the migration of stellar clusters from the disk to the halo. A positive correlation between the number of GCs and the number of mergers was also found in the E-MOSAICS simulation suite \citep{Kruijssen_etal_2019}. Early
observational studies did not find any significant dependence of $\eta$ on
the environment or the morphological type of the galaxy
\citep{Spitler_and_Forbes_2009,Hudson_etal_2014}.  Larger statistical samples
and revised analysis methods have highlighted a second-order difference between
ellipticals and spirals \citep{Harris_etal_2015} in the same sense that is
predicted by our model, but less pronounced. This could suggest a weaker correlation between bulge formation and stellar cluster migration than that assumed in our model.

\begin{figure*}
\centering
\resizebox{16cm}{!}{\includegraphics{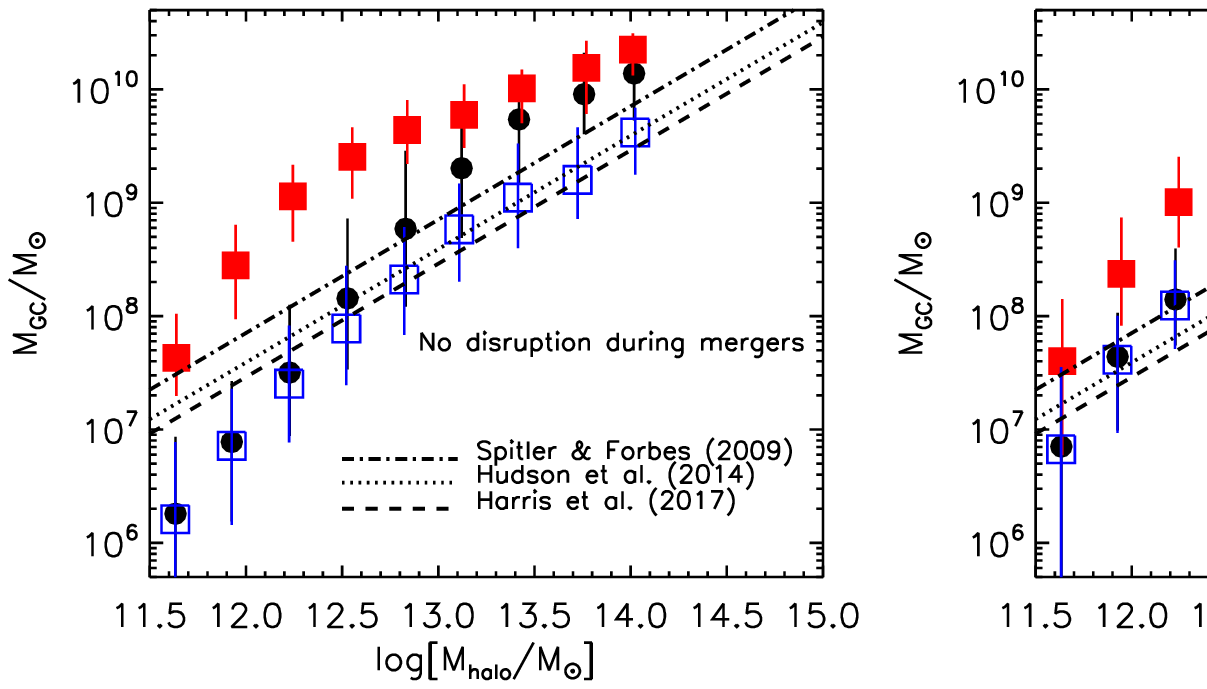}} 
\caption{As in the left panel of Figure~\ref{fig:MgcMhalo} but for a model in which no efficient cluster disruption during galaxy mergers has been considered (on the left), and for a model where no dynamical friction has been considered (on the right).
\label{fig:MgcMhaloalt}}
\end{figure*}

Analysing our results based on alternative physical prescriptions, we find that
both efficient cluster disruption and a treatment for dynamical friction are
needed in our model to predict both the correct slope and normalisation of the
relation. Figure~\ref{fig:MgcMhaloalt} shows the impact of efficient cluster
disruption during mergers (left panel) and dynamical friction (right panel) on the relation between the total mass in GCs as a function of the parent halo mass. As in
Figure~\ref{fig:MgcMhalo}, we plot the median relation (and percentiles) obtained
for all model galaxies (filled circles), and for different morphological types
(filled and empty squares). Switching off both physical modifications cause an increase of the number (and total mass) of GCs associated with model galaxies. This affects the overal normalisation of the predicted relation. Since
galaxy mergers are, by construction, more important for early-type galaxies,
the lack of an efficient cluster disruption during mergers has a stronger
impact for early-type rather than late-type galaxies, thus further strengthening the morphological dependence and implying that our fiducial model is more consistent with observations. Since early-type galaxies
represent a larger fraction of the overall population with increasing galaxy
stellar mass, these particular physical ingredients also affect the predicted
slope of the relation. Our treatment for dynamical friction reduces the mass of
GCs that is associated with model galaxies, more or less independently of the
galaxy type.  This translates into a significant effect only on the overall
normalisation of the relation, but a negligible effect on its shape.

\begin{figure*}
\centering
\resizebox{16cm}{!}{\includegraphics{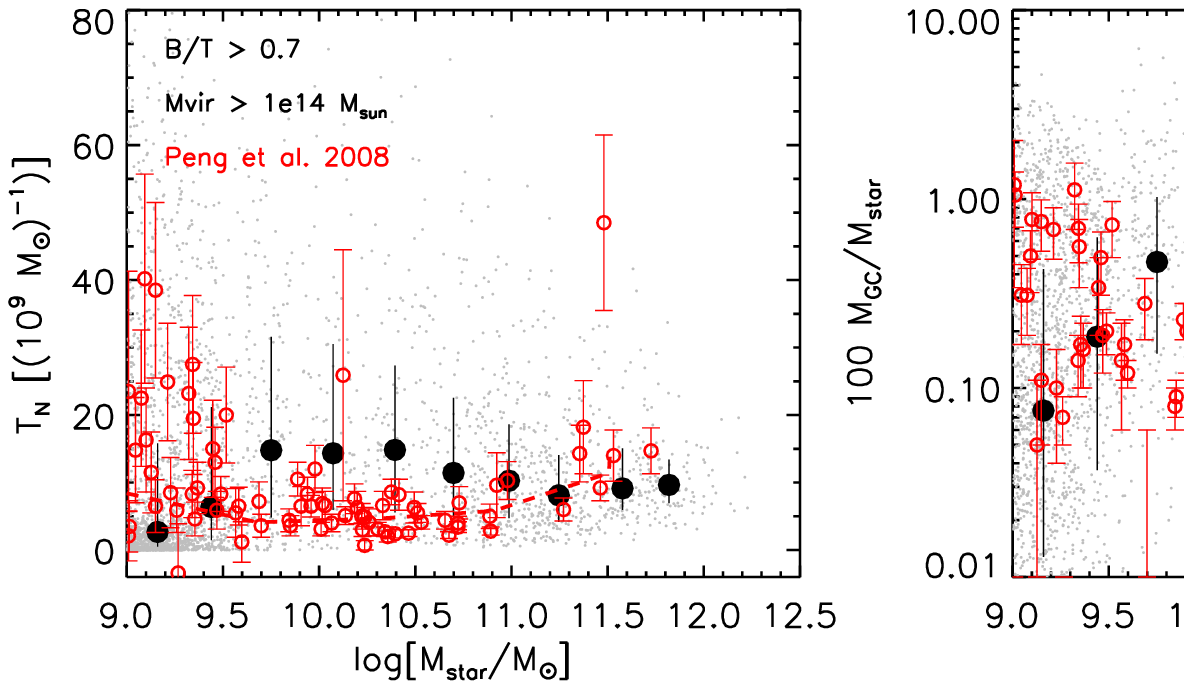}} 
\caption{Left panel: specific frequency of globular clusters as a function of
  galaxy stellar mass. All galaxies with bulge-to-total larger than 0.7 and
  residing in haloes more massive than $10^{14}\,{\rm M}_{\sun}$ have been
  considered in this case, so as to have a fair comparison with observational
  estimates by \citet[][empty red circles]{Peng_etal_2008}. Filled symbols with
  error bars show the median and percentiles (25th and 75th) of the
  distributions of all model galaxies (these are shown as grey dots in the background). Right panel: ratio
  between the total mass in GCs and galaxy stellar mass for the same model
  galaxies considered in the left panel, compared to observational estimates. 
\label{fig:specfq}}
\end{figure*}

Figure~\ref{fig:specfq} shows the predicted dependence of the GCs' specific
frequency, as a function of galaxy stellar mass. For this comparison, we have
considered the same cuts in GC mass, age, and metallicity indicated above, and
we have only considered galaxies (both centrals and satellites) with
bulge-to-total stellar mass ratio larger than 0.7 and residing in haloes more
massive than $10^{14}\,{\rm M}_{\sun}$. These limits have been considered to have a fair comparison with the observational data by \citet[][empty
  symbols with error bars in the figure]{Peng_etal_2008}, based on the Virgo
Cluster Survey conducted with the Hubble Space Telescope. Our simulated sample
considered here consists of about 3000 galaxies, i.e.\ about 40 times larger than
the number of galaxies included in the observational data set over the same
stellar mass range.

\begin{figure*}
\centering
\resizebox{16cm}{!}{\includegraphics{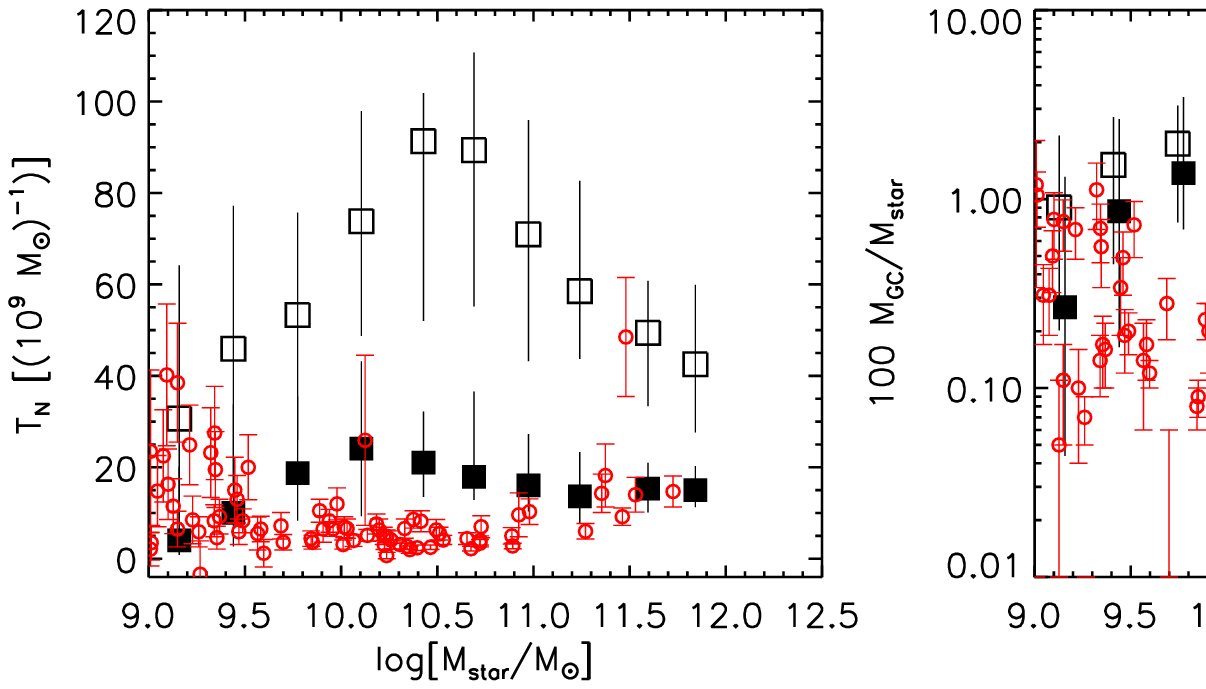}} 
\caption{As in Figure~\ref{fig:specfq}, but showing results for two alternative
  models where no efficient cluster disruption during galaxy mergers (empty
  squares) or no dynamical friction (filled squares) is considered.
\label{fig:specfqalt}}
\end{figure*}

Figure~\ref{fig:specfq} shows that the model predictions cover the same region of observational estimates, with a few noticeable differences. In particular, for the lowest-mass galaxies considered, the observational estimates tend to be larger than our model predictions. The latter exhibits a significant bending of the median GC mass per galaxy stellar mass towards lower values. This might be due to resolution, i.e.\ to the fact that an artificially low number of mergers are resolved in this stellar mass range.  For galaxies with intermediate stellar masses (between $\sim 10^{10}$ and $10^{11}\,{\rm M}_{\sun}$), the median values of our model predictions lie systematically above the observational measurements, with a large number of model galaxies that are characterised by a specific frequency larger than that measured for Virgo galaxies of comparable stellar mass. We note that model predictions shown in Figure~\ref{fig:specfq} correspond to an ensemble of massive haloes. By contrast, the observational data sample is based on one galaxy cluster only. Given the large variations of halo mass accretion histories, and the large expected stochasticity in the merger history of their hosted galaxies, cosmic variance can significantly affect the comparison illustrated in this figure.  

An interesting result shown in
Figure~\ref{fig:specfq}, which is valid in general and not specific to the cluster
environment, is the relatively large scatter of the predicted specific
frequency/mass in GCs at fixed galaxy stellar mass. This result, which can be
appreciated already considering the two specific case studies discussed in
Section~\ref{sec:ex}, is a natural consequence of the stochasticity of the merger
events and of the large scatter of galaxy physical properties and GC population
at the time of mergers. While a general trend as a function of galaxy stellar
mass (see also the next section) exists and is expected (the average number of
galaxy mergers increases as a function of galaxy mass), this scatter is
important and needs to be taken into account to correctly interpret the
observational results.

As discussed above, the mass in GCs (and as a consequence also the specific
frequency of GCs) increases significantly without an efficient cluster
disruption, or when no treatment for dynamical friction is included.
Figure~\ref{fig:specfqalt} shows that, when these two physical processes are
switched off, both the number and the total mass in GCs associated with model
galaxies increase significantly to values that are considerably larger than current
observational estimates. The left panel of the figure shows that the difference
is larger for the specific frequency, suggesting a different shape of the
cluster mass function in these alternative models. We will come back to this in the next section.

%%%%%%%%%%%%%%%%%%%%%%%%%%%%%%%%%%%%%%%%%%%%%%%%%%%%%%%%%%%%%%%%%%%%%%%%%%%%%%%
\section{Mass, metallicity and age distributions of globular clusters}
\label{sec:basicdistr}

In this section, we analyse the physical properties of GCs and their
distributions in terms of mass, metallicity, and age. We will compare these with
available observational measurements for the Milky Way and analyse how the
distributions vary as a function of galaxy stellar mass. For the analysis
presented in this section, we have considered only model galaxies that host at
least one GC. This does not significantly affect the results presented below,
except for the two lowest galaxy stellar mass bins considered. For such low-mass galaxies, the large number of model galaxies with zero GCs (likely due to
resolution limits) has a non-negligible effect on the median distributions
presented.

\begin{figure}
\centering
\resizebox{8cm}{!}{\includegraphics{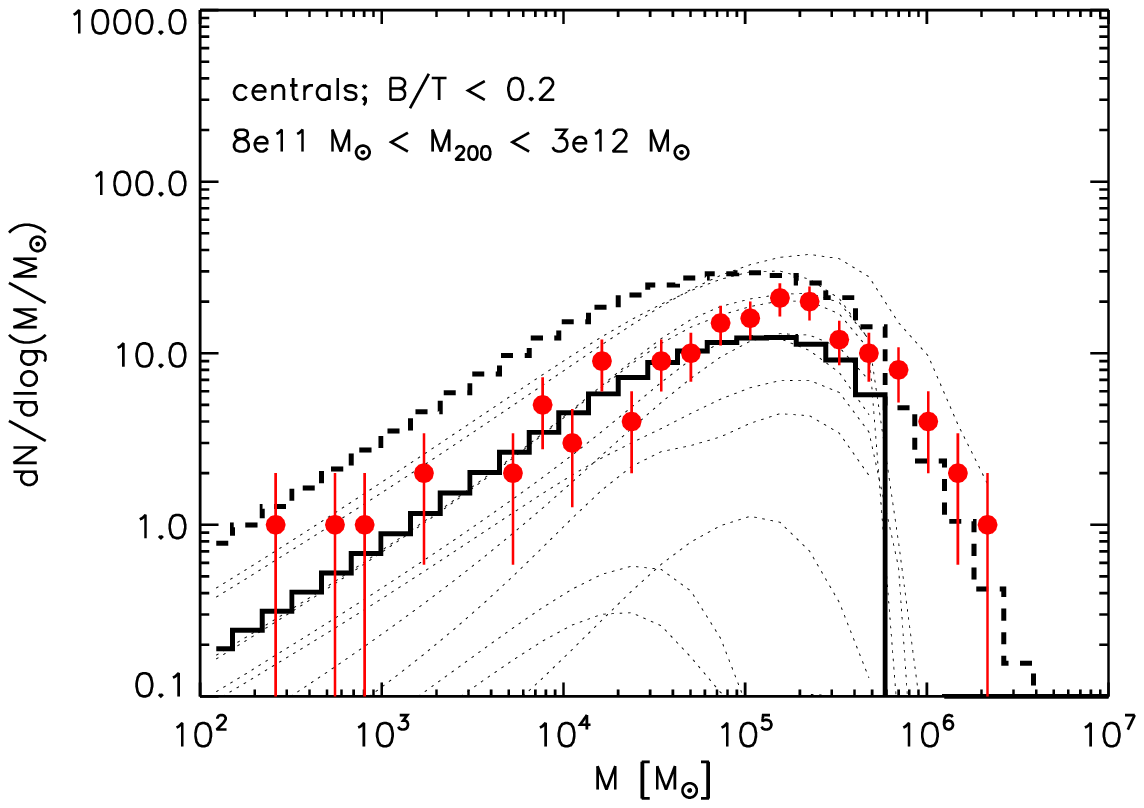}} 
\caption{Mass distribution of globular clusters associated with Milky-Way like
  galaxies. Only model central galaxies of haloes in the mass range indicated
  in the legend have been considered. In addition, only galaxies with a
  bulge-to-total mass ratio smaller than 0.2 (i.e.\ disk dominated galaxies)
  have been selected. The solid  and dashed lines show the median and mean of
  the distribution, respectively. Dotted lines show only a few individual
  `Milky-Way' like galaxies, which have been randomly selected. Filled circles with Poisson error bars correspond
  to observational estimates \citep{Harris_1996}.
\label{fig:mfMW}}
\end{figure}

Figure~\ref{fig:mfMW} shows the mean (dashed line) and median (solid line) mass
distributions of GCs associated with Milky Way-like model galaxies, compared
with the observed GC mass function of the Milky Way (filled circles with error bars; from \citealt{Harris_1996}). The model sample
used for this figure consists of 16,916 galaxies that are disc dominated (the
bulge-to-total mass ratio is smaller than 0.2) and that are centrals of haloes
with mass between $8\times10^{11}$ and $3\times10^{12}\,{\rm M}_{\sun}$.
The dotted lines show a few individual (random) examples. These highlight the
existence of a large galaxy-to-galaxy variance (in terms of normalisation, peak mass, presence of a double peak), which is determined by the different merging histories and physical properties of galaxies at the time of GC formation/migration, even though we consider only a limited halo mass range and galaxies with a low number of mergers (as reflected by the cut in morphological type). For example, when considering individual MW-like galaxies, we find that the peak mass varies between $\sim 10^3\,{\rm M}_{\sun}$ and $\sim 4.5\times10^5\,{\rm M}_{\sun}$.

\begin{figure}
\centering
\resizebox{7.5cm}{!}{\includegraphics{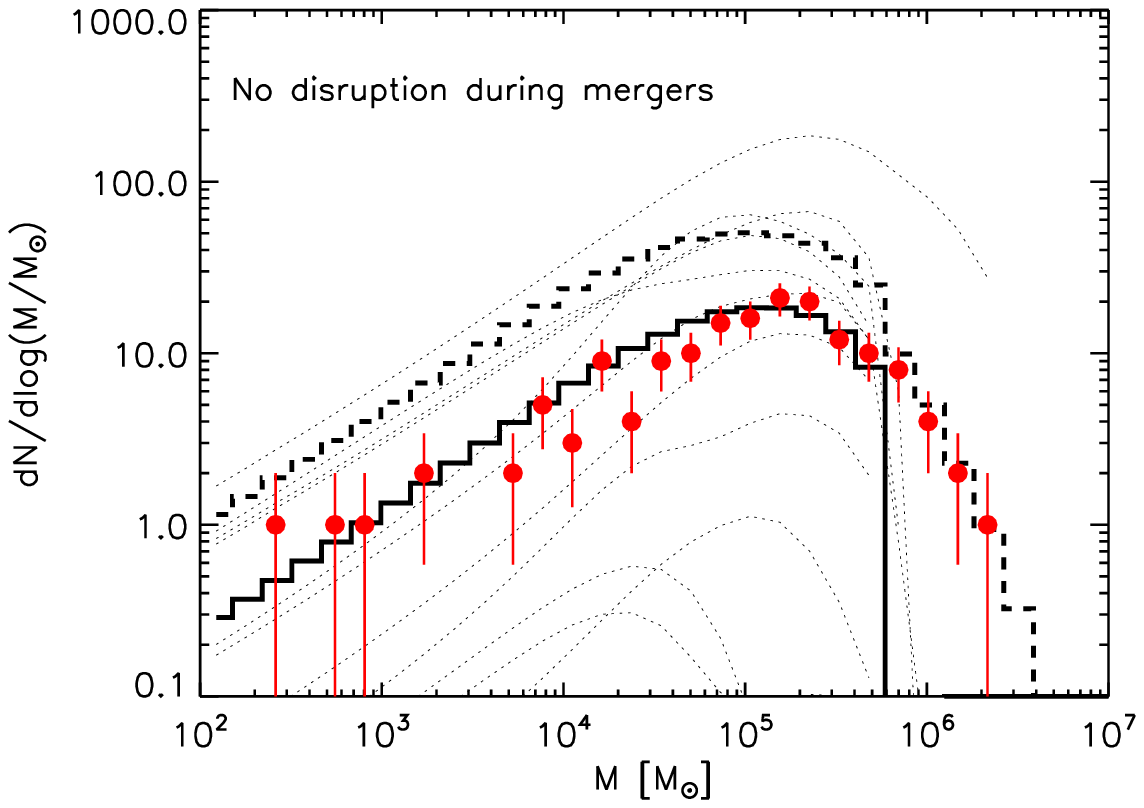}}
\resizebox{7.5cm}{!}{\includegraphics{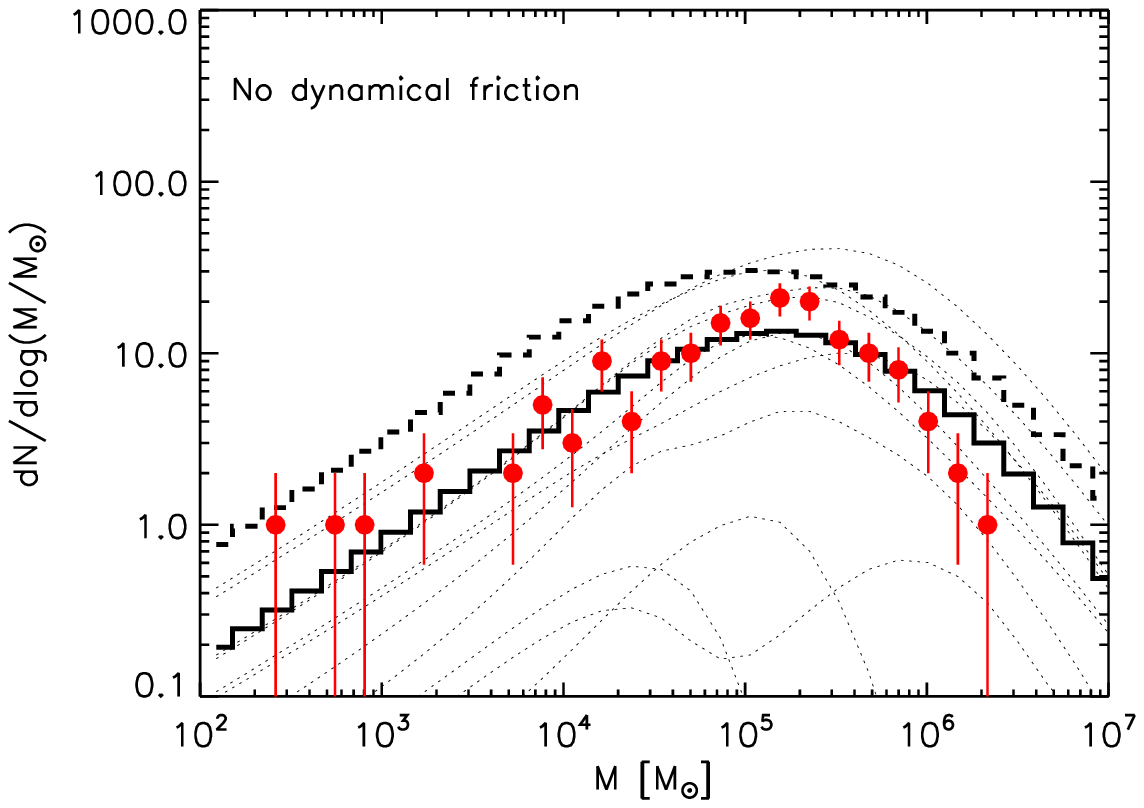}}
\resizebox{7.5cm}{!}{\includegraphics{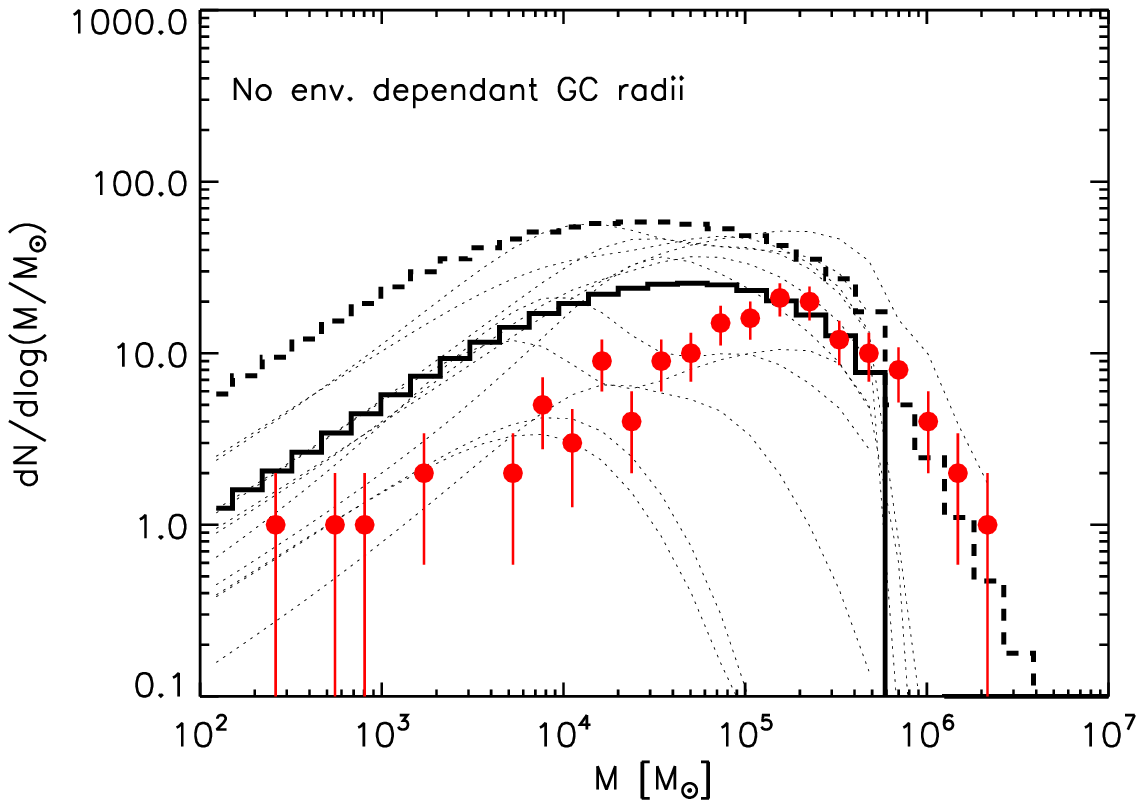}}
\caption{As Figure~\ref{fig:mfMW} but (i) in the top panel no efficient
  cluster disruption in the galaxy disc during galaxy mergers is considered;
  (ii) in the middle panel no model for dynamical friction is considered; (iii)
  in the bottom panel no environmental dependence of cluster radii is considered.}
\label{fig:mfMWalt}
\end{figure}

The median distribution obtained for model galaxies follows the observational
measurements quite well, with a small deficit of GCs at the peak of the
distribution, and a more significant deficit at masses larger than
$\sim6\times10^5\,{\rm M}_{\sun}$ that could be improved by using a less
`aggressive' treatment for dynamical friction (see Section~\ref{sec:dftreat}).
This can be seen in the middle panel of Figure~\ref{fig:mfMWalt}, where we
show the same model predictions, but from a run without a treatment for dynamical
friction. In this run, the predicted GC mass distribution for Milky Way-like
galaxies extends up to $\sim 10^8\,{\rm M}_{\sun}$, with a gradual decrease for
masses larger than $\sim 2\times 10^5\,{\rm M}_{\sun}$. At variance with model
predictions discussed in the previous section, we find that the results shown in
Figure~\ref{fig:mfMW} are not significantly affected if no efficient
cluster disruption during galaxy mergers is considered (see the top panel of
Figure~\ref{fig:mfMWalt}). This is not surprising, given that the sample of
galaxies considered in this case is dominated by late-type galaxies, i.e.\
galaxies that did not experience a significant number of mergers.  Switching
off this physical ingredient actually brings model results in somewhat better
agreement with the observed peak of the mass distribution. However, as we have discussed in the previous section, an efficient cluster disruption during galaxy mergers is needed to reproduce the normalisation of the relation between the total mass in GCs and halo mass. Finally, the bottom panel of Figure~\ref{fig:mfMWalt} shows that, when the assumption of an environmental dependence of cluster radii is relaxed, the peak of the mass distribution is moved to lower masses (for the Milky-Way like galaxies considered here, it moves to $\sim 5\times10^4\,{\rm M}_{\sun}$), and the number of GCs below the peak increases by a factor of $2{-}3$. This is a consequence of the treatment adopted (see Section~\ref{sec:rd} and
Eqs.~\ref{eq:envradfirst}-\ref{eq:envradlast}), which leads to longer survival
times for stellar clusters less massive than $10^4\,{\rm M}_{\sun}$ if a constant, compact radius is adopted.

\begin{figure}
\centering
\resizebox{8cm}{!}{\includegraphics{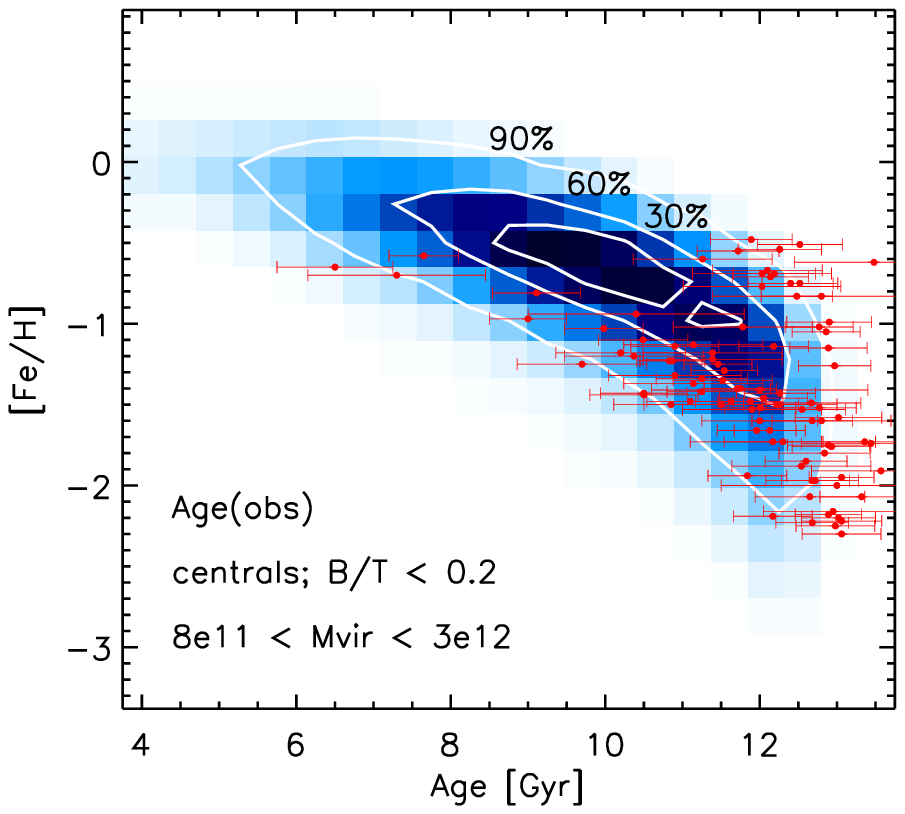}} 
\caption{Age-metallicity distribution of GCs associated with model Milky-Way
  like galaxies. Circles with error bars correspond to individual measurements
  of GCs in our Galaxy (compilation presented in \citealt{Kruijssen_etal_2019b}).
\label{fig:agemetMW}}
\end{figure}

Figure~\ref{fig:agemetMW} shows the age-metallicity distribution of GCs
associated with the same Milky Way-like model galaxies considered above and
compares it with observational estimates of GCs in our Galaxy (circles with
error bars). Considering that observations are cut at [Fe/H]$=-0.5$, the agreement in terms of the metallicity distribution is fairly good. As for ages, our model predictions 
are systematically younger than the observational estimates by $\sim 0.75$~Gyr, on
average. A similar shift was found for the E-MOSAICS simulations \citep{Kruijssen_etal_2019b,Kruijssen_etal_2020}. Specifically, \citet{Kruijssen_2019} found a median value of $\tau_{25}$ (the assembly time of 25~per~cent of the final halo mass) of $10.78$~Gyr, against $11.5$~Gyr estimated for the MW. The same was found for the GCs: the median age of GCs in the MW is 12.26 Gyr, whereas for galaxies in E-MOSAICS the median GC age was found to be $10.73$~Gyr. This shift might be explained by an anomalously early formation time of our Galaxy. In fact, there are several further observations confirming the idea that the Milky-Way is not a `typical' galaxy for its stellar mass, such as the offset from the Tully-Fisher relation \citep{Hammer_etal_2007}, the unusual satellite population \citep[e.g.][]{Geha_etal_2017,Nashimoto_etal_2022}, and the tension between the early formation of the Milky Way's disk inferred from galactic archaeology and state-of-the-art numerical simulations \citep{Semenov_etal_2023}.  In order to verify if this is a generic prediction of the model, or if there are a subset of haloes that would reproduce the data, we have considered only those MW-like model galaxies that have an age distribution `similar' to that measured for our Galaxy. To do so, we have computed, for each model galaxy the Kolmogorov-Smirnov statistics and associated probability that the model age distribution is significantly different from the observed distribution. Considering only galaxies with a low KS probability, we obtain a GC metallicity distribution that is still very similar to the one observed for our Galaxy. However, the resulting GC mass distribution has a significantly different shape and lower normalization than observed, as a consequence of the earlier formation and longer time available for disruption. 

\begin{figure}
\centering
\resizebox{8cm}{!}{\includegraphics{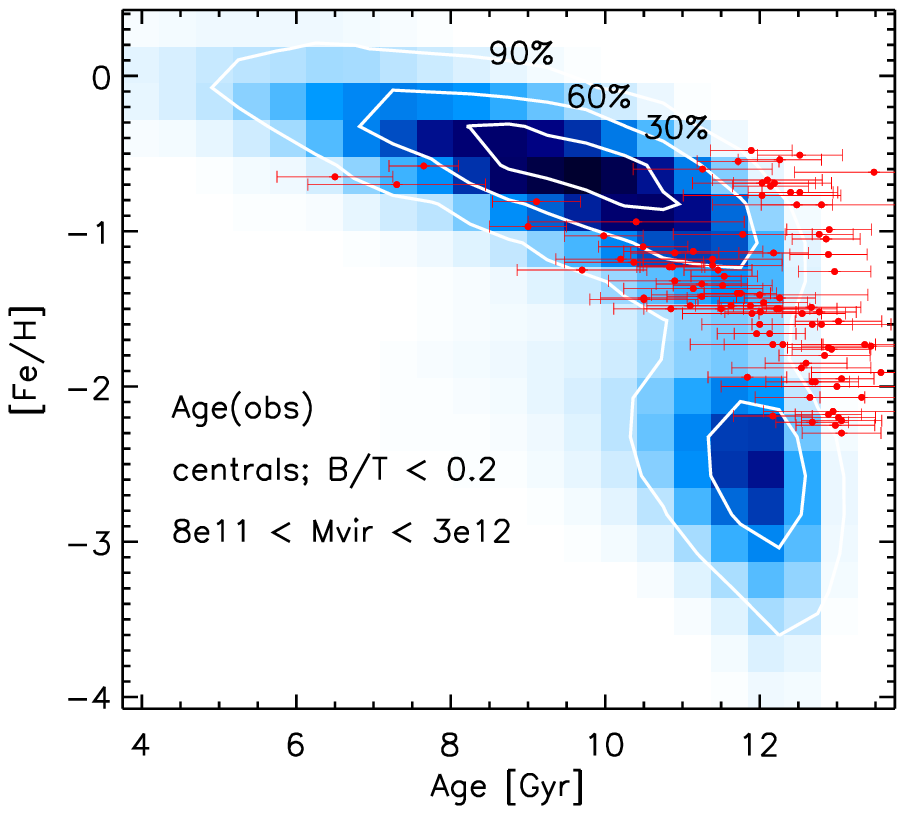}}
\caption{As in Figure~\ref{fig:agemetMW} but for a model in which 95~per~cent of
  the metals in small haloes are ejected directly into the hot gas phase.}
\label{fig:agemetMWmod}
\end{figure}

We have tested the impact of all relevant parameters in the GC model (see Section~\ref{sec:altmodels}), and we find that the only physical process that has a significant impact on the age-metallicity distribution shown in Figure~\ref{fig:agemetMW} is the treatment of chemical
enrichment (more specifically metal ejection) in small haloes.
Figure~\ref{fig:agemetMWmod} shows the age-metallicity distribution of GCs
associated with MW-like galaxies, in a model in which 95~per~cent of the metals
in small haloes are ejected directly into the hot gas phase. The figure shows a
bimodal distribution of metallicities with a second peak at very low values
([Fe/H]$\sim -2.5$). These GCs all have very old ages ($\sim 13$~Gyr) and are
therefore associated with very early episodes of star formation in low-mass
haloes. The `bimodality' that is observed in Figure~\ref{fig:agemetMWmod} could be affected by the limited resolution of the simulation adopted: a higher resolution would resolve a larger number of mergers with lower mass galaxies. This could increase the number (and total mass) of GCs associated with low mass galaxies and affect the metallicity distribution. Although we plan to investigate this in future work, we note here that this feature suggests the prescriptions adopted for chemical enrichment within low-mass haloes may need revisiting. A similar conclusion was also reached independently by studies focusing on the abundances and properties of Damped Ly$\alpha$ absorbers predicted by our GAEA model \citep{DiGioia_etal_2020}.

\begin{figure*}
\centering
\resizebox{16cm}{!}{\includegraphics{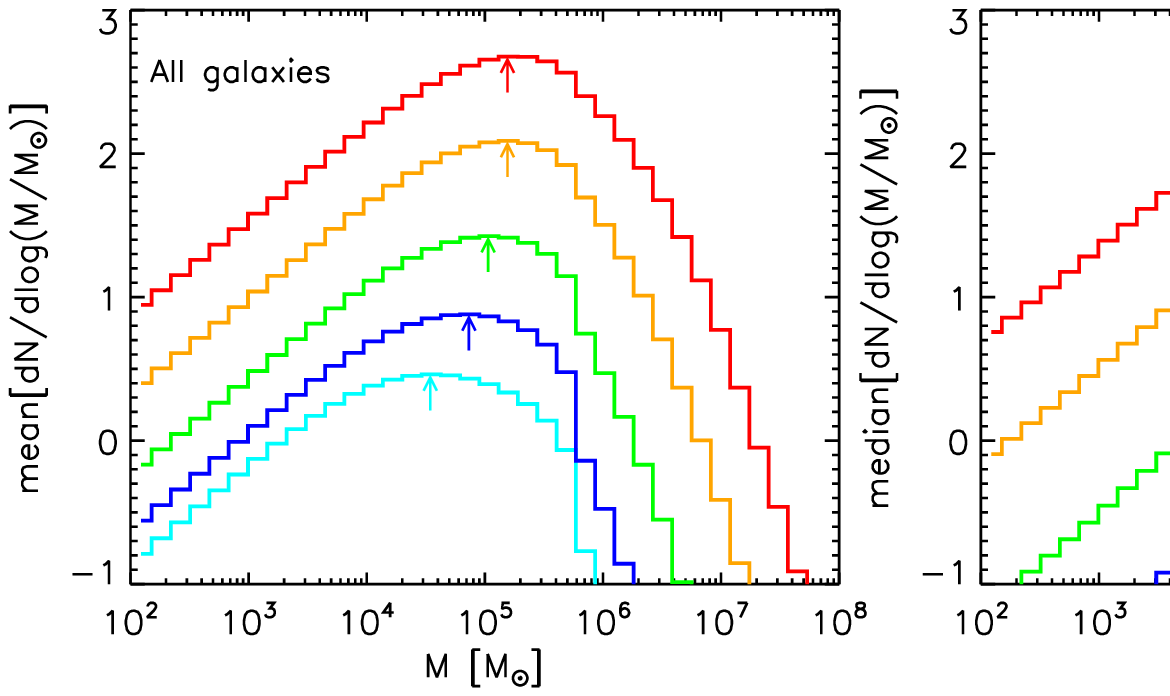}} 
\caption{Mass distribution of globular clusters associated with model galaxies
  of different stellar mass, as indicated in the legend. The left and right panels
  show the mean and median of the distributions, respectively. The vertical
  arrows correspond to the peaks of the corresponding distributions.
\label{fig:mfALL}}
\end{figure*}

Finally, we consider the mass distributions and age-metallicity distributions for model galaxies in bins of galaxy stellar mass, independently of the parent halo mass and morphology.
\begin{figure*}
\centering
\resizebox{17cm}{!}{\includegraphics{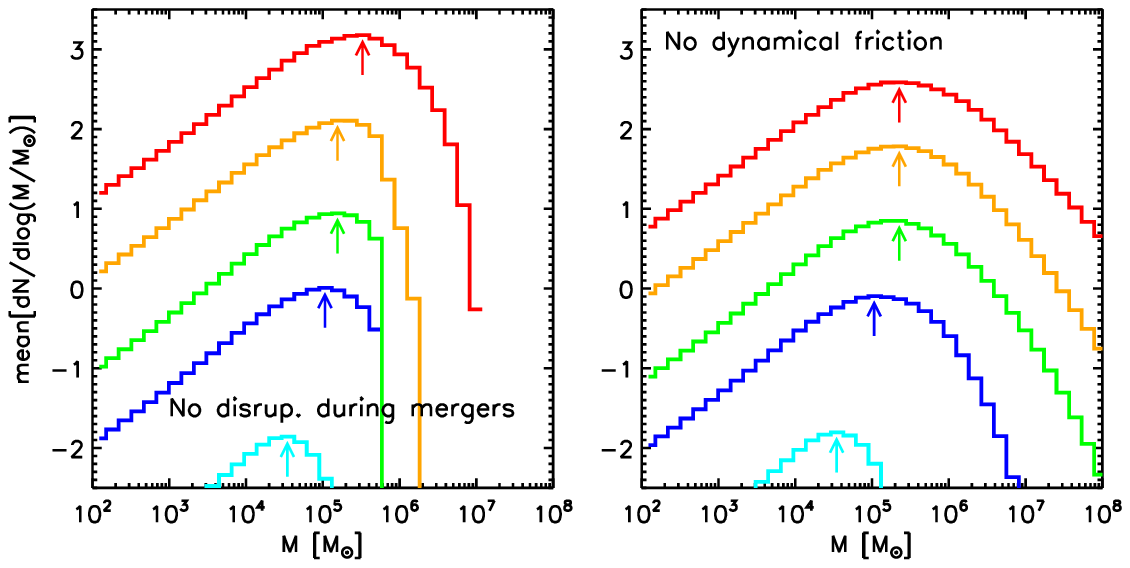}} 
\caption{Median mass distributions of globular clusters associated with model
  galaxies of different galaxy stellar mass (as indicated in the legend), for
  three alternative models considered. The left panel corresponds to a run
  where no efficient cluster disruption in the galaxy disc during mergers is
  considered; the middle panel corresponds to a run with no dynamical friction;
  the right panel corresponds to a run where no environmental dependence of
  cluster radii is assumed.
\label{fig:mfALLmod}}
\end{figure*}
Figure~\ref{fig:mfALL} shows that the peak of the mass distribution increases
weakly with galaxy stellar mass, varying between $\sim 1\times 10^5\,{\rm
  M}_{\sun}$ and $\sim 2\times 10^5\,{\rm M}_{\sun}$ over two orders of
magnitudes in galaxy stellar mass, in qualitative agreement with findings by \citet{Jordan_etal_2007} based on the Virgo Cluster Survey. Figure~\ref{fig:mfALLmod} shows the median mass distribution obtained
in runs where different physical ingredients in our model have been switched
off. The left panel corresponds to a run with no efficient cluster disruption
during mergers; the middle panel to a run with no dynamical friction; the right
panel to a run with no environmental dependence of cluster radii. The impact of
these model prescriptions is consistent with what was discussed for the mass
distribution of GCs in Milky Way-like galaxies. In addition, the figure shows
that the lack of a prescription for dynamical friction virtually removes the
already weak dependence of the peak of the mass distribution on the stellar mass of the galaxy for galaxies more massive than $\sim 10^{10.5}\,{\rm M}_{\sun}$.

\begin{figure*}
\centering
\resizebox{16cm}{!}{\includegraphics{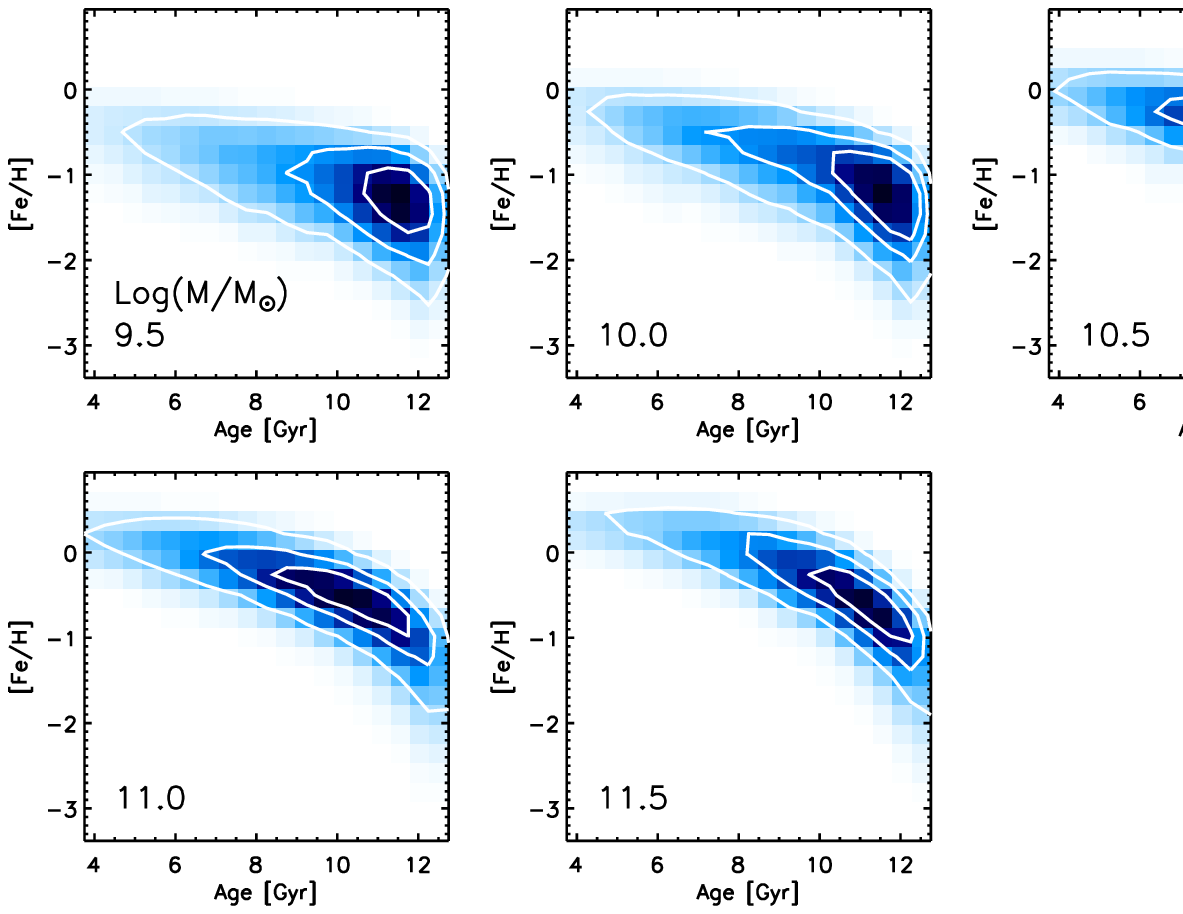}} 
\caption{Age-metallicity distributions of globular clusters associated with
  model galaxies of different stellar mass, as indicated in the legend.
  \label{fig:agemetALL}}
\end{figure*}

Figure~\ref{fig:agemetALL} shows the age-metallicity distribution of GCs
associated with model galaxies of increasing galaxy stellar mass, as indicated
in the different panels. For the lowest mass galaxies considered, most of the
GCs have very old ages and a relatively narrow range of metallicities. As the
galaxy stellar mass increases, both the ranges of ages and metallicities widen,
extending to younger ages and larger metallicities (the two are of course
correlated: younger GCs are formed in gas that has been enriched by previous
generations of stars, and therefore also have larger metallicities).

\begin{figure*}
\centering
\resizebox{16cm}{!}{\includegraphics{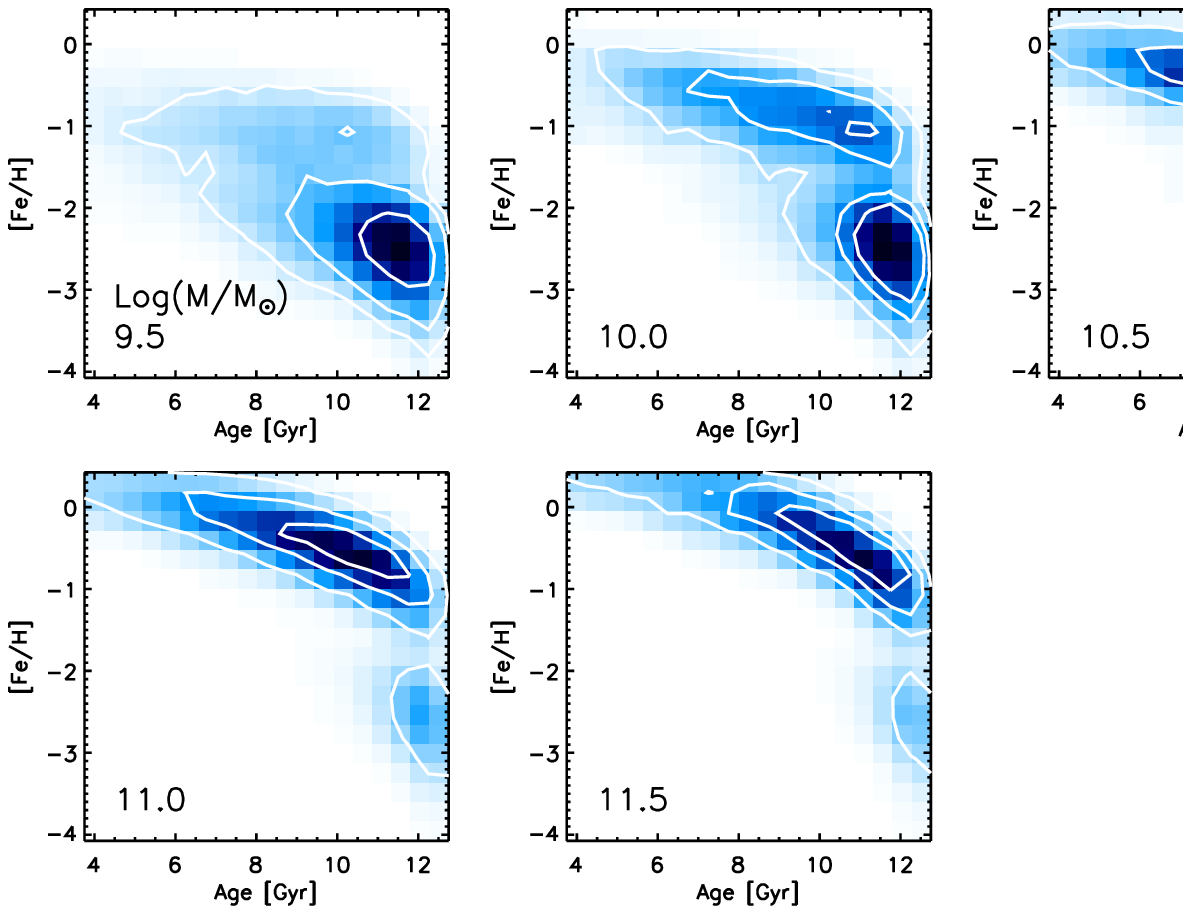}} 
\caption{As in Figure~\ref{fig:agemetALL} but for a model where 95~per~cent of
  the newly synthesized metals in small haloes are ejected directly into the hot gas phase.}
\label{fig:agemetALLmod}
\end{figure*}

Figure~\ref{fig:agemetALLmod} shows the age-metallicity distribution of GCs
associated with galaxies of different stellar mass, for a model where 95~per~cent of the metals in low-mass haloes are ejected directly into the hot gas
phase. The figure shows that this different assumption about metal enrichment
in small haloes leads to the development of secondary density peaks around very
old ages and low metallicities, for all galaxy masses considered. This secondary peak at very low metallicities and old ages is dominant in the least massive galaxies considered, supporting the warning given above that this result might be affected by resolution.

%%%%%%%%%%%%%%%%%%%%%%%%%%%%%%%%%%%%%%%%%%%%%%%%%%%%%%%%%%%%%%%%%%%%%%%%%%%%%%%
\section{Discussion and Conclusions}
\label{sec:discconcl}

In this paper, we present an end-to-end description of the formation process of
GCs, that combines a treatment for their formation and dynamical evolution
within galaxy haloes with a modelling of the latter through a state-of-the-art
semi-analytic model of galaxy formation and evolution. This theoretical framework has been constructed by effectively coupling the GC model presented in \citet{Kruijssen_2015} and the GAlaxy Evolution and Assembly (GAEA) semi-analytic model \citep{DeLucia_etal_2014, Hirschmann_etal_2016}. A similar approach has
been recently used in the framework of the E-MOSAICS project
\citep{Pfeffer_etal_2018,Kruijssen_etal_2019}.  Being based on hydrodynamical 
simulations, however, these studies are limited to relatively small cosmological
boxes ($\sim 50$~Mpc on a side) or to a few individual re-simulations. Our
fully semi-analytic approach requires significantly shorter computational
times, allowing us to:

\begin{itemize}
  \item[(i)] efficiently explore the coupling between the galaxy and star cluster
    formation physics parameter space by running a large number of model
    variants;
  \item[(ii)] model large populations of galaxies to study the effect of
    environment and assembly history with exquisite statistics;
\end{itemize}

Our approach also improves upon previously published models that parametrize
the formation of the GC populations using dark matter accretion histories
extracted from cosmological simulations \citep[e.g.][]{Kruijssen_2015,Choksi_etal_2018,El-Badry_etal_2019}, by including an explicit treatment for galaxy formation and 
the effect of the evolving galactic environment on star cluster formation and 
disruption. 

Our model reproduces naturally the observed correlation between the total mass
in GCs and the parent halo mass \citep{Spitler_and_Forbes_2009,Harris_etal_2017}.
The predicted relation is linear for haloes more massive than $\sim 3\times10^{12}\,{\rm M}_{\sun}$, with a deviation from linearity for lower halo masses. In the framework of our model, such a deviation is not affected by GC formation physics, at least not by the processes that we have explicitly varied: a fixed/varying value for the minimum cluster mass scale, the environmental dependence of GC radii, the efficiency of cluster disruption during mergers, and dynamical friction. In our model, the turnover at low masses is driven by a significant dependence on morphological type in this mass range: bulge dominated galaxies host, on average, larger masses of GCs than their late-type counterparts resulting in a closer to linear behaviour at low halo masses. This dependence  is a natural consequence of our assumption that cluster migration from the disk to the halo is triggered by galaxy mergers, and that bulges are predominantly built through mergers. Although a similar dependence is seen in the observational data, the observed effect is not as strong as predicted by our model, which might suggest a weaker correlation between bulge formation and the migration of stellar clusters than predicted by our model. 

One caveat to bear in mind is that haloes with mass $\sim 10^{12}\,{\rm M}_{\sun}$ are resolved with $\sim 800$ particles in the Millennium Simulation that we have employed in our study. This might result in merger trees that are not resolved well enough to give convergent results for the corresponding central galaxies: increased resolution would lead to a larger number of mergers (with smaller haloes) and these could potentially bring in a larger number of GCs from accreted galaxies. This could reduce the bending of the predicted relation at low halo masses, and potentially also the different mass of GCs predicted for the corresponding central galaxies that have late and early-type galaxies. Such a difference is, however, predicted also for galaxies that are well resolved within the simulation employed in this study, and therefore represents a robust prediction of our model. 

We find that both the slope and the normalization of the predicted relation between halo mass and total mass in GCs depend on cluster physics. In particular, our model requires both an efficient cluster disruption during galaxy mergers and a rather aggressive treatment for dynamical friction to bring model predictions in close agreement with observational results: the former affect both the slope and the normalization of the relation because of the stronger impact on early-type galaxies and the varying fraction of these galaxies as a function of galaxy stellar mass; the latter has a significant effect on the overall normalization but a negligible impact on the slope. 

Our reference model reproduces quite well the observed mass distribution of GCs in our Galaxy, with a deficit at GC masses larger than $6\times10^5\,{\rm M}_{\sun}$ that could be improved with a less aggressive treatment for dynamical friction. At lower GC masses, our model requires an environmental dependence of GC radii (i.e. lower survival times for less massive clusters) to bring model predictions in agreement with observational data. The model GCs in Milky-Way galaxies tend to be systematically younger (by $\sim 1$ Gyr) than those observed in our Galaxy. A similar result was found in the E-MOSAICS simulations \citep{Kruijssen_etal_2019b,Kruijssen_etal_2020} and could be explained by an anomalously early formation of our Galaxy. The metallicity distribution is unimodal, with a peak at [Fe/H]$\sim -0.5$. When considering the overall galaxy population, our model also predicts a weak increase of the peak of the GC mass distribution with increasing galaxy stellar mass, in qualitative agreement with observational results. The predicted age-metallicity distribution depends significantly on galaxy mass: both the age and metallicity ranges widen for more massive galaxies, that include younger and more metal rich GCs. Again, there is no clear sign of bimodality in either the age of metallicity distributions obtained when considering the entire galaxy population in the simulated volume. 

As mentioned above, the large volume of the simulation translates into large statistical samples of galaxy populations. Our sample of Milky-Way like galaxies, selected only on the basis of halo mass and bulge-to-total mass ratio and using only about 10~per~cent of the Millennium Simulation volume, is made up of about 60,0000 galaxies. Our results highlight the existence of a very large galaxy-to-galaxy variance, even in the limited halo mass bin of Milky-Way like haloes, that is driven by the different galaxy merger histories and physical conditions at GC formation and migration. When considering individual galaxies, bimodal or multi-modal metallicity and age distributions become not uncommon. In future studies, we plan to investigate in further detail how the predicted properties of GCs depend on the mass accretion and merger history of their parent galaxy/halo.

%% If GCs are relatively old and
%% the z = 0 GC population is viewed as the composite population of
%% GCs formed in progenitor halos and assembled through mergers, no
%% coupling between GCs and dark matter halos is needed to explain
%% the observed relation, at least at high halo masses

%%%%%%%%%%%%%%%%%%%%%%%%%%%%%%%%%%%%%%%%%%%%%%%%%%%%%%%%%%%%%%%%%%%%%%%%%%%%%%%
\section*{Acknowledgements}
GDL gratefully acknowledges support from the Alexander von Humboldt Foundation, and the hospitality of Heidelberg University, where part of this work was carried out.
JMDK gratefully acknowledges funding from the DFG through an Emmy Noether Research Group (grant number KR4801/1-1).
STG gratefully acknowledges the generous and invaluable support of the Klaus Tschira Foundation.
JMDK and STG gratefully acknowledge funding from the European Research Council (ERC) under the European Union's Horizon 2020 research and innovation programme via the ERC Starting Grant MUSTANG (grant agreement number 714907).
COOL Research DAO is a Decentralised Autonomous Organisation supporting research in astrophysics aimed at uncovering our cosmic origins.
MH acknowledges funding from the Swiss National Science Foundation (SNF) via the PRIMA Grant PR00P2 193577 “From cosmic dawn to high noon: the role of black holes for young galaxies”.

\section*{Data Availability}

The model data underlying this article will be shared on request to the corresponding author. An introduction to GAEA, a list of our recent work, as
well as data files containing published model predictions, can be found at: https://sites.google.com/inaf.it/gaea/ 

\bibliographystyle{mnras}
\bibliography{sc_sam} % if your bibtex file is called example.bib

%%%%%%%%%%%%%%%%%%%%%%%%%%%%%%%%%%%%%%%%%%%%%%%%%%%%%%%%%%%%%%%%%%%%%%%%%%%%%%%

% Don't change these lines
\bsp	% typesetting comment
\label{lastpage}
\end{document}